\documentclass{article}

\usepackage{geometry}
\usepackage{graphicx}
\usepackage{newtxtext}
\usepackage{newtxmath}
\usepackage[square,sort,comma,numbers]{natbib}
\usepackage{hyperref}

\usepackage{amsmath}

\newcommand{\rood}[1]{}

\usepackage[T1]{fontenc} 
\usepackage[utf8]{inputenc}

\usepackage{xcolor}
\usepackage[ruled,vlined]{algorithm2e}

\SetCommentSty{mycommfont}
\usepackage{bm}
\usepackage{cases}
\graphicspath{{././}}
\usepackage[section]{placeins}
\usepackage{siunitx}

\hypersetup{
    colorlinks = true,
    urlcolor   = blue,
    citecolor  = black,
}

\newcommand{\RomanNumeralCaps}[1]
\linenumbers

\DeclareUnicodeCharacter{2212}{\ensuremath{-}}

\makeatletter							
\def\printtitle{
	{\centering \huge \sc \textbf{\@title}\par}}		
\makeatother

\makeatletter							
\def\printauthor{
	{\centering \small \@author}}				
\makeatother

\title{Morphological transitions of block copolymer micelles: implications for isoporous membrane ordering}

\author{%
 Nicolas Moreno$^{1*}$, Suzana Nunes$^{2}$, and Victor Calo$^3$ \\
{1.Basque Center for Applied Mathematics (BCAM), Alameda de Mazarredo 14, Bilbao 48400, Spain\\
 2. Biological and Environmental Science and Engineering Division, Advanced Membranes and Porous Materials Center, King Abdullah University of Science and Technology (KAUST), 23955-6900, Thuwal, Saudi Arabia\\
 3. School of Electrical Engineering, Computing and Mathematical Sciences, Curtin University, Kent Street, Bentley, Perth, WA, 6102, Australia\\
 {$*$ nmoreno@bcamath.org} }

 }
\date{2023}

\begin{document}
\printtitle 

\printauthor

\begin{abstract}
Isoporous membranes made from diblock copolymers have numerous applications, including water treatment and protein separation, and are successfully produced at a laboratory scale under controlled conditions. However, achieving optimal conditions for membrane preparation remains a challenge due to the complexity of the involved phenomena. Experimental studies have shown that the self-assembly of block copolymers in solution significantly affects the final membrane structure. Despite extensive research, understanding the multiscale phenomena leading to the characteristic morphology is still elusive. We address this gap by using mesoscale computational simulations to investigate the self-assembly of block copolymers in selective solvents, consistent with isoporous membrane preparation. We focus on the interplay between entropic and enthalpic interactions and their effects on the morphology of the micellar aggregates in solution. Our computational results are consistent with experimental evidence, revealing a morphological transition of the aggregates as the polymer concentration and solvent affinity change. We propose different phase parameters to characterize the emergence of monodisperse-spherical micelles in solution and describe the order of crew-cut micelles using a rigid-sphere approximation. Our study provides valuable insights into the self-assembly of diblock copolymers to optimize the preparation of isoporous membranes.
\end{abstract}

\section{Introduction}

Highly ordered isoporous membranes can be manufactured from diblock copolymers self-assembly via non-solvent induced phase separation~\cite{ Nunes2010, Nunes2011, Dorin2012, Marques2013, Rangou2013, Yu2015}. The membrane fabrication is a multiscale process that involves different stages. The process starts with the solubilization of the block copolymer in a mixture of selective solvents, followed by the casting of the polymeric solution, after which part of the solvent evaporates. Finally, the solution film is immersed in water to induce the complete phase separation creating the final membrane. The mechanism of pore formation in these membranes combines block copolymer self-assembly in solution and macrophase separation by solvent exchange. When films are prepared by complete solvent evaporation, the structures present in the original solution are progressively lost and approximate the equilibrium bulk morphology. The advantage of the membrane formation by phase inversion is that the structures in the casting solution are immobilized to give rise to the characteristic pores.

Micellar order in solution is crucial to fabricating isoporous membranes with nano-scale resolution. Different investigations suggest that the arrangement of the micelles in solution directly affects the final configuration of the pores of the membranes~\cite{ Nunes2010, Nunes2011, Phillip2011, Dorin2012, Oss-Ronen2012, Marques2013, Marques2014}. Thus, the type of order in the solution resembles the final membrane~\cite{ Dorin2012, Oss-Ronen2012, Marques2013}. Hexagonal or body-centred cubic (BBC) like arrangements of micelles in solution produced the equivalent inverted structure (i.e., hexagonal-like porous) in the fabricated membrane.

Since the ordering of micelles in solution is a prerequisite for the isoporous formation, identifying the parameter space (i.e., polymer concentration ($w$), block copolymer composition ($f_A$), and polymer-polymer and polymer-solvent interactions) where ordered states occur, offers a step forward in the understanding and controlling of the membrane production.  At high polymer concentrations, long-range ordering of diblock-copolymer micelles in solution has been reported and rationalized~\cite{ McConnell1997}. In those conditions, the relative length ratio of the block is responsible for BCC to face-centred cubic (FCC) order transitions, while the polydispersity of the diblock copolymer has important effects in order-disorder transition~\cite{ Lynd2007}. In the context of isoporous membranes, unfortunately, the governing features of micellar order in solution remain unclear. 

Here, we rationalize micellar assembly stemming from geometrical constraints and use mesoscale simulations to estimate the phase diagram of the assembled morphologies. Based on these geometrical considerations, we study the effect of the polymer concentration, solvent-polymer interactions and molecular weight of the block copolymer. In the following sections, we first define the polymeric systems of study and formulate the geometrical conditions that determine the ordering in solution. Next, we describe the computational model construction. Finally, we propose a set of phase parameters that describe the dependence of the phase diagram on the system's different entropic and enthalpic variables.

\begin{figure}[t!]
\centering
\includegraphics[trim=0cm 0cm 0cm 0cm, clip=true,width=0.8\textwidth]{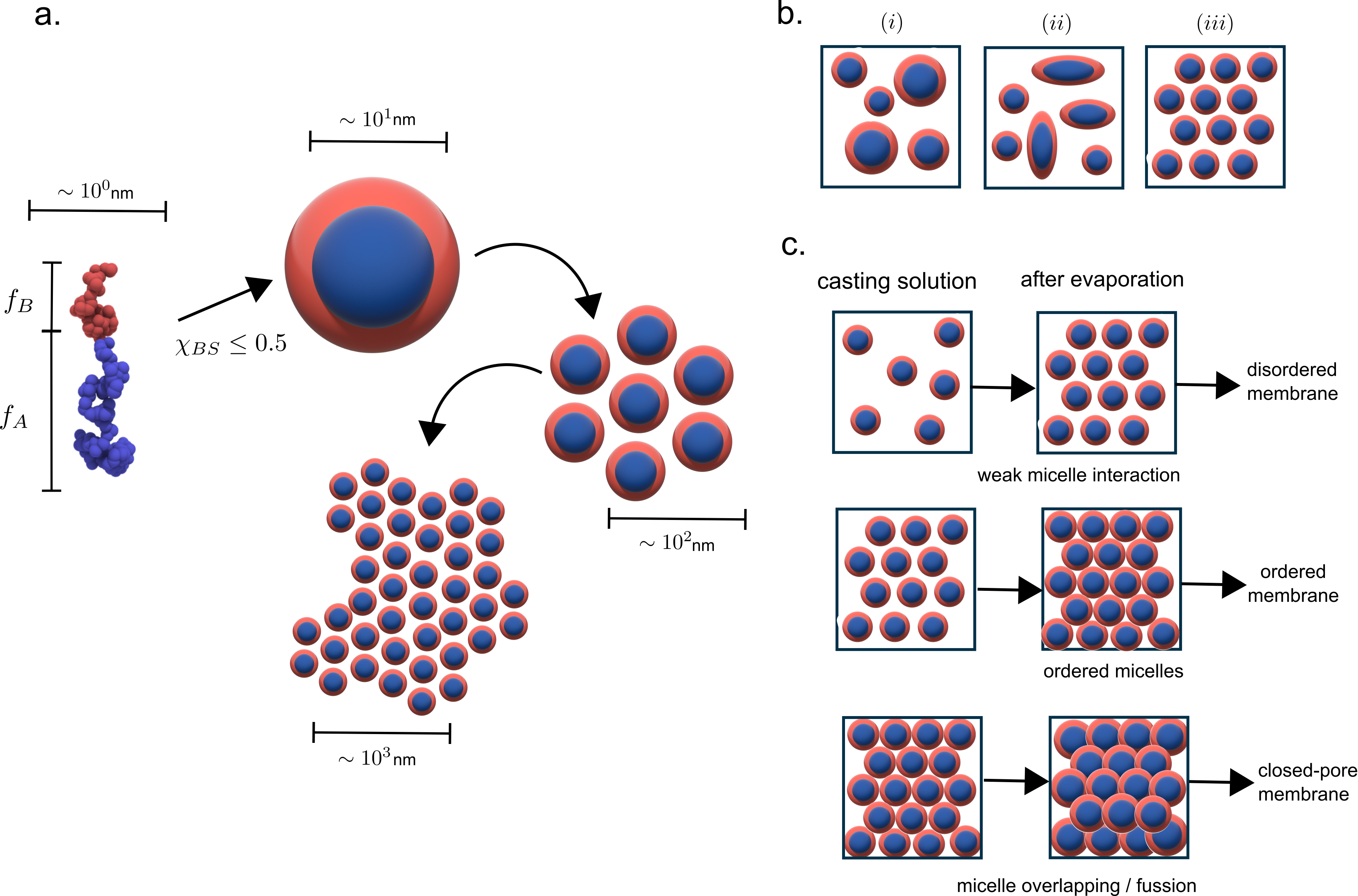}
\caption{a) Sketch of hierarchical self-assembly of block copolymers in solution. b) Crew-cut micelle illustration for different block compositions and solvent selectivity. Variations in thermodynamic conditions lead to various micelle morphologies. c) Different scenarios can describe the micelles concentration effect on the ordering of micelles in casting solutions after the evaporation stage.
}
\label{fig:bldblocks}
\end{figure}

\section{Hierarchical Self-Assembly of Block Copolymers}

We can describe the self-assembly of block copolymers in the bulk or melts by considering the relative volume fraction of each \textit{i}-th component ($f_i$), the degree of polymerization of the block copolymer ($N$), and the Flory-Huggins interaction parameters~\cite{ Flory1942} ($\chi_{ij}$) between \textit{i}-th and \textit{j}-th component. In the case of the semidilute polymer solutions, Lyotropic liquid crystalline states may occur and additional concentration-dependent parameters are needed~\cite{ Forster1996}. The formation of mesoscale structures with long-range ordering in the casting solution for isoporous membranes entails a hierarchical assembly of the block copolymers into sub-units (i.e., micelles) that constitute the building blocks of the mesoscale aggregates, as Figure~\ref{fig:bldblocks}.a illustrates. Some factors such as the relative length ratio of the corona and core~\cite{ McConnell1997, McConnell1995}, and the polydispersity of the diblock copolymer~\cite{ Lynd2007} have important effects in the order-disorder transition. 

Knowing the system's free energy is possible to analyze the block copolymers' self-assembly and the conditions where ordered micelles occur. However, due to the multi-component nature of the mixture and the nonlinearity of the polymeric solution with the concentration, there is no univocal free-energy expression and a variety of formulations exist elsewhere in the literature. These free energy expressions mostly focus on the micellization process in dilute systems~\cite{ Nagarajan1989, Whitmore1990, Shi1995, Lin1996, Nagarajan1999, LaRue2008, Hadgiivanova2011, Jensen2012, Nagarajan2014}, and only a few studies have considered the effects of micelle-micelle interactions~\cite{ McConnell1996}. Scaling theories, widely used to describe the bulk phase behaviour of block copolymers, give us an alternative to understanding the hierarchical assembly. 
Different scaling laws in the literature describe the assembly of crew-cut and star-like micelles; these scaling regimes (power law relationships) describe the mesoscopic phases' size depending on the blocks' relative volume fraction and $\chi_{AB}$. In particular, we expect that geometrical constraints affect the $N_i$-scaling of each microdomain (i.e., A and B). These scaling approaches express the size ($R \propto \gamma^{\delta} N_A^{\alpha}N_B^{\beta}$) and aggregation number ($Z \propto N_A^{\Gamma}N_B^{\Lambda}$) of the micelles in terms of the total degree of polymerization and the core-corona interfacial tension, $\gamma$. The magnitude of the exponents $\delta$, $\alpha$, $\beta$, $\Gamma$, and $\Lambda$ is characteristic for each micelle type. These scaling theories only capture the effect of the block copolymer molecular weight under fixed polymer-solvent interactions. Scaling parameters for different segregation strengths exist, but their explicit dependence with $\chi_{ij}$ remains elusive. For crew-cut micelles, for example, these scaling relations~\cite{ Karayianni2016} exist
\begin{align}
R_C \approx \gamma^{1/3} N_A^{2/3} v_A^{1/3}&&
Z \approx \gamma N_A.    
\end{align}
This scaling model assumes uniformly stretched chains in the core and a good solvent for the corona-forming block. Under these conditions, the aggregation number of the micelles is only a function of the degree of polymerization of the core-forming block, $N_A$. In the case of block copolymers in solution, scaling relations analogous to those for bulk are applicable after minor corrections~\cite{ Forster1996} to describe the aggregation number of amphiphilic block copolymers as $Z \approx Z_0 N_B^{-0.8}N_A^2$, where $Z_0$ accounts for the packing parameter of the block copolymer. In semidilute block copolymers solutions, where long-range lattice structures emerge, Birshtein and coworkers~\cite{ Birshtein1990} proposed that scale laws for the size of the shell and core do not depend on the concentration with the exponent ${2/3}$.	
%

\section{Geometric Description of Self-Assembly in Solution}

We define the casting solution for the membrane preparation as a semidilute solution of an amphiphilic diblock copolymer $AB$ in a solvent $S$ (or a mixture of solvents) selective for the $B$ block ($\chi_{AS} > 0.5 \geq \chi_{BS}$). Under these circumstances, the block copolymers form aggregates that are microscopically phase-separated into $A-$ and $B-$rich domains.  $ N_i$ segments constitute each $i$th block, and the total number of segments per chain is $N = N_A + N_B$. We express the composition of the A block as $f_A = N_A/N$. Typically membranes are prepared with asymmetric block copolymers ($N_A>N_B$), which favor the formation of core-shell micelles~\cite{ Avogadro}. Depending on the system's energy balance the solvent molecules can be found in the core and shell of the micelles. We denote the volume fraction of the polymer and solvent in the solution as $w$ and $w_S = 1 - w$, respectively. For simplicity, the volume of the segments, $v_i$, is assumed to be the same for both blocks and for the solvent that is represented as one segment ($v_A = v_B = v_S$). The degree of polymerization, the Khun length, of a particular polymer blob can characterize the size of the segments and the polymer chain. Here, we associate the segments with the mesoscale beads used to construct the polymer chains. In particular, we use the dissipative particle dynamics method~\cite{ Groot1997, Search1992, Search1995}.

In a casting solution with a total volume $V$, containing $M$ micelles, the number density of micelles is $\rho_M = M/V$. We assume that $\left\langle Z \right\rangle$ block copolymer chains constitute each micelle in equilibrium with solvent particles, where $\left\langle Z \right\rangle$ is the average aggregation number, $Z$, per micelle. In the following, for simplicity, we omit $\left\langle \cdot \right\rangle$, when referring to properties of the micelles. However, they always refer to the assemble average unless otherwise stated.  Regarding the block copolymer and solvent molecules in the system, the sample volume is $V = (N_Av_A + N_Bv_B) + N_Sv_S$, where $N_S$ is the total number of solvent particles. Similarly, after the assembly of the polymer chains, $V$ can be recast in terms of the volume of the formed micelles as
\begin{equation}
V = V_M M + lv_S,
\end{equation}
where $V_M$ is the average volume of the micelles, and $l$ is the number of free solvent molecules not directly associated with the block copolymer within the micelles. For aggregates with core-shell morphologies $V_M = V_{core}+V_{shell}$, being $V_{core}$ and $V_{shell}$ the volume of the core and shell of the micelle, respectively. Since the micelles are in equilibrium with the solvent, we denote the amount of solvent localised in the micelles' core and shell as $j$ and $k$, respectively. Thus, the total number of solvent particles is $N_s = l+j+k$.

By taking into account the block ratios in the copolymer ($f_A$ and $f_B$) and the selectivity of the solvent, we expect aggregates to exhibit crew-cut morphologies~\cite{ Zhang1996a, Zhang1999}, characterized by the larger and less solvent-selective block $A$, forming a dense core, while the short copolymer $B$ forms the shell (or corona) of the assembly~\cite{ Nunes2011} (see Figure~\ref{fig:bldblocks}.a). We construct the phase diagram for our systems for different affinities between the blocks ($\chi_{AB}$) and solvent selectivity ($\chi_{AS},\ \chi_{BS}$) to validate the crew-cut morphologies. We verify that the most stable aggregates correspond to crew-cut micelles under relevant membrane preparation conditions.

\subsection{Packing of Rigid Micelles}

The packing of micelles in a solution can be analyzed geometrically by considering the micelles as polymer-solvent pseudo-phases in equilibrium. The characteristic packing of the micelles depends on their shape, size distribution, and concentration in the system. In Figure~\ref{fig:bldblocks}.b, we present three possible micelle shape and size distribution scenarios, each affecting the packing in distinct ways. Case \textit{(i)} in Figure~\ref{fig:bldblocks}.b corresponds to a system with polydisperse spherical micelles in equilibrium, where ordering in solution is entropically prohibited due to the significant difference in size between the assembled micelles. Therefore, hexagonal close-packed (HCP) or BCC lattices reported~\cite{ Nunes2011, Dorin2012} cannot be stable. For case \textit{(ii)} in Figure~\ref{fig:bldblocks}.b, the solution contains monodisperse non-spherical micelles, which also impose entropic constraints that impede micellar arrangement in hexagonal-close-packed configurations. Finally, for case \textit{(iii)} in Figure~\ref{fig:bldblocks}.b, monodisperse spherical micelles assemble in the solution. In this case, closely-packed configurations can be achieved if the concentration of micelles is appropriate.

We rationalize the ordering of the micelles in the case \textit{iii} on Figure~\ref{fig:bldblocks}.b as a packing of rigid spheres. Carl Friedrich Gauss demonstrated that  for close-packed configurations (i.e., HCP and face-centred cubic (FCC)) the highest volume fraction of monodisperse-rigid spheres ($w_{max}$) is approximately $0.74$. Therefore, if HCP or FCC arrangements are formed with monodisperse spheres, the micellar volume fraction in the system, $w_M$, and the volume fraction of free solvent ($w_S^o = 1 - w_M$) should be close to $w_{max}=0.74$. We use this volumetric constraint to define the transitions between aggregation regimes of our system. For instance, volumetric fractions larger than $w_{max}$ lead to the fusion of micelles in a block copolymer solution containing equilibrated monodisperse micelles of radius $R_M$. The micelle density ($\rho_M$) in the system, without overlapping, is given by $\rho_M = {w_{max}}/{V_M}$, where $V_M = 4/3 \pi R_M^3$, is the average volume of a micelle. We treat the micelles as rigid spheres, and $w_{max}$ approximates the maximum micellar volume fraction before fusion occurs. In real systems, the corona of the micelles can overlap without fusing the core of the micelles; therefore, the value of $w_{max}$ that we adopt can be slightly smaller than those in real experimental conditions. Nevertheless, using this rigid-sphere approximation allows us to identify the ordering transitions. Assuming spherical micelles, the volume of the core and shell domains are given by
\begin{equation}
V_c = \frac{4}{3}\pi R_c^3 = v_AN_AZ + jv_s,
\end{equation}
and,
\begin{equation}
V_s = \frac{4}{3}\pi ((R_c+R_s)^3-R_c^3) = v_BN_BZ + kv_s,
\end{equation}
where $R_c$ and $R_s$  are the core and shell radius, respectively. The amount of solvent in equilibrium with the micelles is given by the variables $j$ and $k$. 			

Identifying the critical condition for micelle overlap allows us to find the polymer concentration regime for isoporous membrane preparation. When polymer solutions are cast for membrane preparation, part of the solvent evaporates, slightly increasing the polymer concentration. The evaporation time determines the extent of the polymer concentration increase. However, due to the short time used during the evaporation, the polymer does not reach equilibrium as the solvent composition changes; instead, the morphological changes that may occur are kinetically governed by the former equilibrium state of the casting solution.        
Regarding the polymer concentration effect, we analyze three different situations. First, when the volumetric fraction of the micelles is much smaller than the close packing condition $w_M \ll w_{max}$, the ordering in solution is not stable, leading to weakly-interacting micelles; under these circumstances, the ordering of the micelles cannot be preserved after the non-solvent phase separation, and disordered membranes are produced. The second regime occurs when the micelle volume fraction is closer but still smaller than $w_{max}$ ($w_M \lesssim w_{max}$). In the proximity of $w_{max}$, the micelles are close to their order condition, such that after evaporation, the micelles remain organized, interacting strongly. This strong association of the micelles warrants preserving the particular pore distribution in the membrane when immersed in water. The third concentration regime occurs when $w_M > w_{max}$. The larger values of $w_M$ lead to the fusion of the micelles modifying their packing; moreover, as the evaporation takes place, the micelles completely overlap, and the ordering of the membrane cannot be sustained. In Figure~\ref{fig:bldblocks}.c we illustrate the effect of micellar density over the ordering in solution and the expected order in the final membrane. 

The scenarios proposed to describe the effect of the casting polymer concentration have been confirmed experimentally~\cite{ Marques2013}. 
For membranes prepared from different initial block copolymer concentrations (PS-P4VP) in a solvent mixture. For this system, initial polymer concentrations smaller than 16\% are suitable for isoporous membrane preparation. After the evaporation, the increase in polymer concentration is sufficient to pack the micelles closely without breaking the ordering. In contrast, casting polymer concentrations larger than 17\% cannot sustain the packing of the micelles after evaporation. Thus, before water immersion, the lamellar structure is formed due to the fusion of the micelles. Thus, the pores are no longer organized in hexagonal order and rather elongated and larger holes appear.


\section{Methods}

Considering the previous geometric constraints and system description, we use the dissipative particle dynamics (DPD)~\cite{Search1992, Search1995} method to conduct the mesoscale simulations. DPD has been widely used to simulate complex mesoscale fluids and soft matter systems. For example, DPD is popular in polymer science to study the self-assembly of block copolymers~\cite{ Marques2013, Yu2015a, Xie2016b, Moreno2015e, Moreno2020, Nikkhah2023}, as it allows for the investigation of large lengths and time scales, which are not feasible with molecular simulations. Additionally, DPD can model solvent-mediated interactions, making it useful for simulating the behaviour of polymers in solution. DPD can incorporate the effects of polymer-solvent interactions and has been shown to reproduce conformational coil features of polymer coils in solution~\cite{ Moreno2014b, Moreno2015a}. Recent applications of DPD include the study of the dynamics of polymers in confined geometries~\cite{Yang2013}, the self-assembly of block copolymers and surfactants in solution~\cite{ Groot1997, Moreno2015e, Xiang2019, Wang2022, Panoukidou2019, Scacchi2023}, nanoparticles~\cite{ Chen2011g, Ma2018, Moreno2020}, gels~\cite{ Buglakov2022} and emulsions. DPD has also been used to study biological systems, such as the dynamics of lipid bilayers~\cite{ Gao2007}, proteins~\cite{ Moreno2015}, and the behaviour of red blood cells~\cite{ Moreno2013, Pan2011}.

In DPD simulations, the fluid is represented as particles interacting with each other and their surroundings. The interactions are modelled based on dissipative forces, which represent energy dissipation due to viscosity and thermal fluctuations, and random forces, which represent the thermal fluctuations of the particles. In DPD, the kinematic evolution and the balance of linear momentum of the particles are given by
\begin{align} 
\frac{d\textbf{r}_i}{dt} &=\textbf{v}_i,
\\
m_i\frac{d\textbf{v}_i}{dt} &=\textbf{f}_i = \sum_{j \neq i}(\textbf{F}_{ij}^C+\textbf{F}_{ij}^D+\textbf{F}_{ij}^R),
\end{align}
where $\textbf{r}_i$, $\textbf{v}_i$ are the position and velocity of a particle $i$, respectively, $m_i$ is its mass, and $\textbf{f}_i$ is the net force acting on the particle. The force acting on each particle has three different contributions; $ \textbf{F}_{ij}^C$ is a conservative force that models pressure effects between particles and spring interactions in chain models. $\textbf{F}_{ij}^D$, models dissipative (viscous) interactions
in a fluid (a friction force that reduces the velocity differences between particles). $\textbf{F}_{ij}^R$ is a random force (stochastic) that models random collisions between particles. This stochastic force approximates the Brownian motion of polymers and colloids. From the statistical mechanics' point of view, $\textbf{F}_{ij}^D$ and $\textbf{F}_{ij}^R$ are tightly related to satisfying the fluctuation-dissipation theorem, which takes the  form of the Fokker-Plank equation~\cite{ Search1995}. The conservative force can be written as $\textbf{F}_{ij}^C = \textbf{F}_{ij}^{B} + \textbf{F}_{ij}^{S}$, where $\textbf{F}_{ij}^{B}$ and $\textbf{F}_{ij}^{S}$ account for bead-bead and bead-spring (when particles are connected) interactions, respectively~\cite{ Posel2014}. In terms of their energy potentials $u_{ij}$, the bead-bead and bead-spring contributions can be expressed as
\begin{align}
\textbf{F}_{ij}^B &= -\frac{\text{d}u_{ij}^B}{\text{dr}_{ij}}\frac{\textbf{r}_{ij}}{|r_{ij}|}, 
\\
\textbf{F}_{ij}^S &=   -\delta_{ij}\frac{\text{d}u_{ij}^S}{\text{dr}_{ij}}\frac{\textbf{r}_{ij}}{|r_{ij}|},
\end{align}
where $\textbf{r}_{ij} = \textbf{r}_i - \textbf{r}_j$ and $r_{ij} = |\textbf{r}_{ij}|$.  $\delta_{ij} = 1$ if particles $i$ and $j$ are connected, and $\delta_{ij} = 0$ otherwise. In the literature, the most used bead-spring energy potentials are harmonic and finite-extensible-non-linear elastic springs~\cite{ Symeonidis2005, Posel2014}, however, other alternatives are possible~\cite{ Symeonidis2005}. Regarding the bead-bead contribution, soft-repulsive potentials are typically chosen as the simplest option~\cite{ Groot1997}, nevertheless, other potentials can be used~\cite{ Pagonabarraga2001, Soddemann2003}. We adopt a classical bead-bead potential~\cite{ Backer2005, Fuchslin2009} and a harmonic spring potential to model the bead-spring interactions such that
\begin{align}
u_{ij}^B &= \frac{a_{ij}}{2r_c}(r_{ij}-r_c)^2,
\label{eq:bbPotential}
\\
u_{ij}^S &= \frac{K_s}{2}(r_{ij}-r_o)^2, 
\label{eq:bsPotential}
\end{align}
where $a_{ij}$ is an interaction (repulsion) parameter, $K_s$ is the spring constant and $r_o$ corresponds to the equilibrium average distance between particles.  The remaining forces are defined as
\begin{align}
\textbf{F}_{ij}^D&=-\gamma \omega ^D(r_{ij})\left(\frac{\textbf{r}_{ij}}{|r_{ij}|}\cdot
\textbf{v}_{ij}\right)\frac{\textbf{r}_{ij}}{|r_{ij}|},
\\
\textbf{F}_{ij}^R&=\sigma \omega ^R(r_{ij}) \zeta \Delta t^{-1/2}
\frac{\textbf{r}_{ij}}{|r_{ij}|},
\end{align} 
where $\gamma$ is a friction coefficient that determines the overall magnitude of the dissipative term, $\sigma$ is the noise amplitude that scales the stochastic contribution, $\omega ^D$ and $\omega ^R$  are weighting functions that set the range of interaction between particles, $\zeta$ is a random number with zero mean and unit variance. The different forces satisfy Newton's third law and conserve linear and angular momenta.  According to Espa\~nol and Warren~\cite{Search1995},  from the fluctuation-dissipation theorem, the dissipative and stochastic force are coupled, and require $\omega^D(r_{ij}) = [\omega^R(r_{ij})]^2$ and $\sigma^2 = 2\gamma k_BT$, where $k_B$ is the Boltzmann constant and $T$ is the equilibrium temperature. The weighting function $\omega^R(r_{ij})$ (and therefore $\omega^D(r_{ij})$) is given by $\omega^D(r_{ij}) =[\omega^R(r_{ij})]^2= (1-r_{ij}/r_c)^2$ for $(r_{ij}<r_c)$, and $\omega^D(r_{ij})=0$ for $(r_{ij}\geq r_c)$. 

We use LAMMPS~\cite{Plimpton1995} to conduct the simulations. The systems were evolved for $1e10^6$ time steps, with a time step 0.04$\tau$ to control temperature fluctuations. We define the simulations units of length, $r_c = 1$, mass $m = 1$, energy $\epsilon = k_BT =1$, time $\tau = 1$ and particle density $\rho_n = 3$ particles$/r_c^3$, according to Groot and Warren~\cite{Groot1997}. We use cubic domains of size $L_{box}=131r_c$, with periodic boundary conditions. The box size is 13-fold  the unperturbed radius of gyration, $R^o_g$, such that the characteristic size of the ordered structures is commensurable in this box size. We use a bead-spring constant $K'_s = 50 k_BT$ with an equilibrium length of $r_o = 0.8$. The visualization and analysis of the results in this document were carried out with in-house codes and the software OVITO~\cite{Stukowski2010, Stukowski2013}.

The simulations have two stages: \textit{i)} an initialization stage where the block copolymer is mixed with the desired solvent, and \textit{ii)} the production stage where the systems reach the equilibrium morphology at this polymer-solvent condition. We start the simulations considering the block copolymer is initially in an athermal solvent for both blocks ($\chi_{ij} = \chi_{ii}$), and the system is equilibrated for $1\text{x}10^5$ time steps. Then the interaction parameters between the less soluble block and the solvent are gradually increased during $5\text{x}10^5$ time steps, until the desired solvent condition. After this initialization stage, all the simulations are conducted for $1\text{x}10^6$ time steps. We use the initialization step in our simulations to simulate the experimental preparation of the casting solution, which is typically prepared in time scale on the order to hours.

The Flory-Huggins interaction parameters give the relative affinity of the solvent with each block, $\chi$~\cite{ Flory1942}, which defines the interaction parameters, $a_{ij}$,  of the DPD method. One alternative to use the Flory-Huggins parameters to map qualitatively the solvent affinity is to verify that the size of the polymer coil follows the characteristic scaling law $R_g \propto N^{\nu}$~\cite{ Rubinstein2003}, where the polymer-solvent affinity governs the magnitude of $\nu$. Values of $\chi \approx 0.5$ ($\nu \approx 1/2$) are expected to represent theta-solvent, $\chi \gg 0.5$ ($\nu \approx 1/3$) poor solvent, and $\chi < 0.5$ ($\nu \approx 3/5$) good solvent~\cite{ Rubinstein2003}. The computed $\chi$ helps define the solvent quality, but the diblock copolymer's final phase separation (e.g., micelle formation) is governed by the magnitude $\chi N$.  We define the interaction parameters of the A block ($\chi_{AS}$) with the solvent and the interaction between blocks ($\chi_{AB}$), according to the polymer coil conformational transitions~\cite{ Moreno2015}, and compute an effective $\left(\chi_{ij}N\right)_{eff}$ for the DPD simulations, following the mapping proposed by Groot et al~\cite{ Groot1998}.
\begin{equation}
\left(\chi_{ij}N\right)_{eff} = \frac{0.306(a_{ij}-a_{ii})N}{1.+3.9N^{-0.51}},
\end{equation}
where $a_{ii} = 25.0$. We adopt the interval of $a_{ij}$ proposed by Moreno et al.~\cite{ Moreno2015a}, that delimits the transitions between good ($a_{ij} < 27.5$), athermal, theta ($a_{ij} \approx 27.5$), and poor ($a_{ij} > 27.5$) solvents. This interval is valid under the length and spring potentials we use to construct the copolymers. The methodology allows us to construct DPD models that mimic the coil size and conformation of the blocks if the relative affinity between the mixture of solvents and the constituent blocks of the system is known. We construct block copolymers with a resolution of $N=192$ beads per chain to characterise the coil conformation transitions with the interaction parameters. The block composition of the block copolymer is fixed in $f_A = N_A/N = 0.75$. 

Investigating micellar ordering in solution would typically require simulations-box size on the order of $\mathcal{O} \gg R_g \approx N^{\nu}$.  Thus, the coils resolution ($N=192$) used to characterize the conformation of the polymers imposes practical limitations to explore the phase behaviour of this polymeric system thoroughly due to its computational cost. Here, we use the fine-scale simulations ($N=192$) to characterize the scaling ($\nu$) of each block and adopt a polymer-model-reduction methodology~\cite{ Moreno2015a} to build tractable polymeric systems.

The conformation of the fine-scale ($\nu'_i$) and the number of beads ($N_i$) of each block ($i$) geometrically restrict the coarse-graining level ($\phi = N_i|_{\text{fine scale}}/N_i|_{\text{coarse scale}}$) that can be applied while keeping the conformational features of the polymer coil~\cite{Moreno2015a}. Since the maximum level ($\phi_{max}$) of reduction is different for the A and B blocks due to their size disparity, we adopt the smallest of the two blocks, which in this case corresponds to the B block. We use a level of coarse-graining in all the simulations $\phi = 14.5$ (Appendix summarises the scaling functions use to construct the coarse polymer models). For all the coarse-grained coils, the scaled DPD potential incorporates a harmonic bond-angle potential with a constant $K_{a,ref} = 9 k_bT$.

\section{Results}

\begin{figure}[!htb]
\centering
\includegraphics[width=0.9\textwidth]{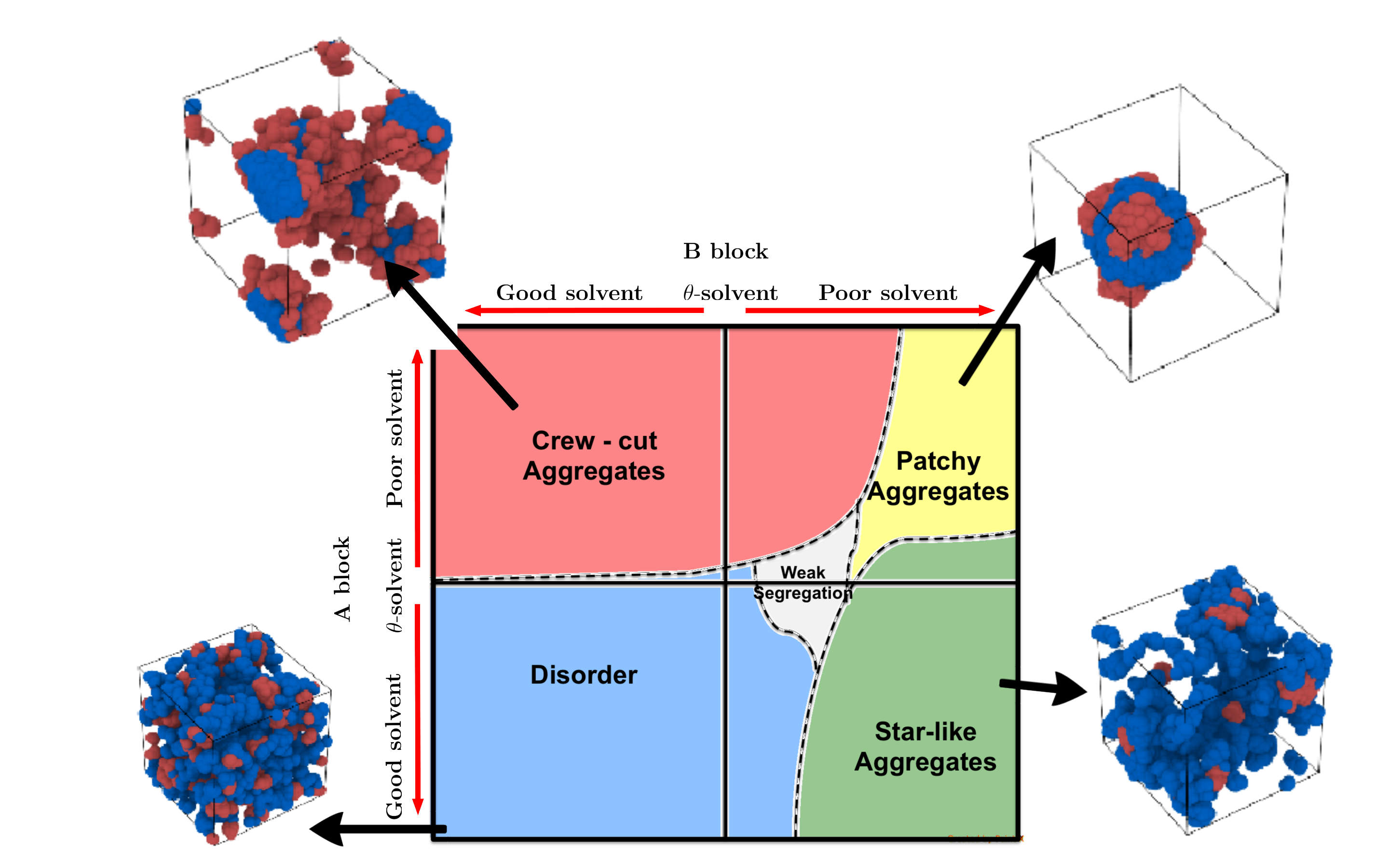}
\caption[Phase diagram for micelle transitions for different polymer-solvent interactions]{Morphological transition of micelles for different polymer-solvent interactions. Blue particles correspond to block~A, and pink to block~B. Solvent particles are omitted to facilitate the visualization. Crew-cut and star-like micelles can be stabilized under the proper solvent conditions. In poor-solvent conditions for both blocks, patchy aggregates emerge. Transitions between phases are sketched based on the simulation results and do not correspond to the exact solvent interactions where the morphology changes.}
\label{fig:blockLocalization}
\end{figure}

The balance of the interaction parameters between the species in the solution dictates the microphase separation of the block copolymer. Thus, the orientation of the blocks inside the micelles is expected to depend on the relative magnitude of $\chi_{ij}$. We verify the block copolymers assembly into crew-cut micelles by evaluating various solvents, varying the interaction parameters for each block ($\chi_{AS}$ and $\chi_{BS}$) from poor to good solvent conditions. We also consider four block copolymers with different levels of affinity between their blocks ($\chi_{AB}$). Figure~\ref{fig:blockLocalization} presents the phase diagram for weak block-block interactions ($\chi_{AB} =0.9$). We corroborate that crew-cut morphologies are favorable over star-like aggregates if the solvent is selective to the shortest block ($\chi_{AS} > \chi_{BS}$). We identify that even if the repulsion between the A and B blocks is weak, the A block localizes in the inner part of the assemblies due to its large size. For athermal A-B interactions ($\chi_{AA} = \chi_{BB} = \chi_{AB}$), there is no microphase separation, and both blocks disperse uniformly in the polymer blob. The balance of pair-wise interactions between species determines the most favorable block localization in the assemblies. Block localization changes can be compared with the energy balance in a triple-point contact A-B-S, where the surface tension between species gives the contact angle; see Figure~\ref{fig:blockLocalization}.

\subsection{Phase Diagrams of Micelle Assembly}

Given that crew-cut micelles appear under the casting-solution conditions, we study the effect of the polymer-solvent interactions and polymer concentration on the shape and size distribution of the assemblies in block copolymer solutions. We aim to identify the parameter space that favors monodisperse and spherical micelles. We use the geometrical-packing interpretation to analyze the phase transitions. Using the entropic ($f_A$, $N$, $w_i$) and enthalpic ($\chi_{ij}$) contributions, we propose the phase parameters that describe the experimental conditions for monodisperse-spherical micelles in a non-dimensional form. In analogy to order-disorder phase diagrams for block copolymers in bulk, we propose a phase diagram for solutions of block copolymers to predict conditions order, disorder and assembly morphologies.

\begin{figure}[!htb]
\centering
\includegraphics[width=1.\textwidth]{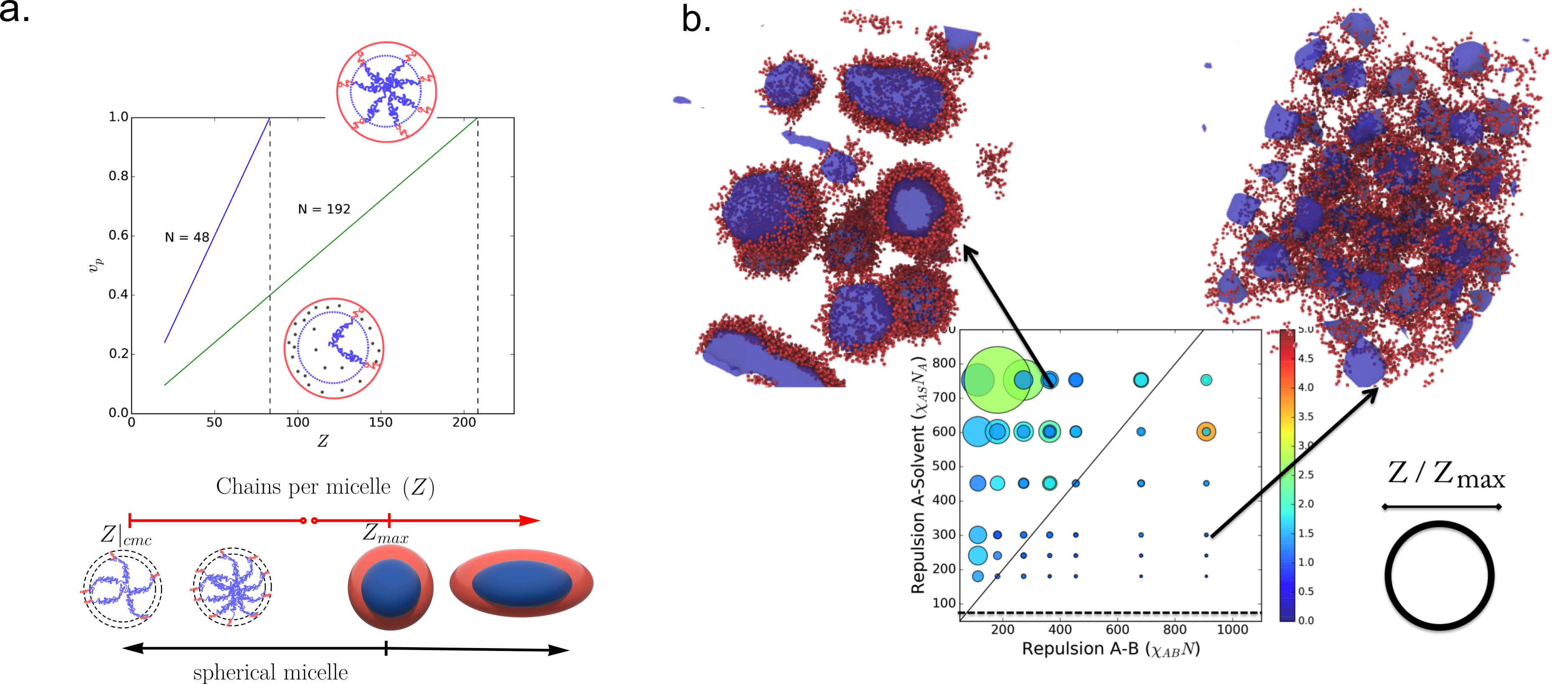}
\caption[Phase diagram description]{a.) Polymer composition of spherical micelles as the aggregation number increases ($v_p  = (NZ/\rho_M)/V_M$). Dashed lines indicate the magnitude to $Z$ when block copolymer chains only form the micelles. b.) Phase diagram description. In the diagram, the axis represents the interactions between the A block and solvent, $\chi_{AS}$, compared to the block interactions $\chi_{AB}$. The interactions of the corona-forming block and the solvent $\chi_{BS}$ are fixed. The color of the markers indicates the polydispersity index of the aggregates, and the size is given by the ratio $Z/Z_{max}$. The dashed line represents the order-disorder transition (ODT). This transition indicates the critical A-solvent interactions where the self-assembly of the block copolymer chains occurs. The regions formed by block~A are rendered as surfaces, whereas the B blocks in the corona are depicted using particle representation. Solvent particles are omitted} 
\label{fig:PDexample}
\end{figure}

We characterize the aggregation state of the simulated systems by computing the micelle-number density,  the coil size of each block, the average aggregation number, and the size of the micelles. Since our goal is to identify the parameter space that  favors monodisperse and spherical micelles, we use the polydispersity index of the assemblies (PdI) to quantify the size distribution. In contrast, the aggregation number $Z$ indicates the sphericity of the micelles. The PdI is given by the number average ($Mn$) and weight average ($Mw$) molecular weight of the assemblies as $\text{PdI} = Mw/Mn$. We account for the spherical nature of the micelles by normalizing $Z$ with the maximum aggregation number of a spherical micelle with the same volume containing only block copolymer ($Z_{max}$).  Figure~\ref{fig:PDexample}.a shows the variation of the polymer composition of spherical micelles of size $R_M$, as a function of the aggregation number for two block copolymers with different $N$. The maximum aggregation number at constant volume corresponds to pure-block copolymer spherical micelles (no solvent inside, i.e., $j=k=0$). Further increase in the aggregation number is only feasible if the shape of the micelles is no longer spherical.  

We streamline the presentation of our results by using phase diagram representations as Figure~\ref{fig:PDexample}.b illustrates. This representation allows us to show the variation of both PdI and sphericity of the micelles. In the diagrams, the color of the markers indicates the PdI of the system, whereas the size of the markers corresponds to the ratio $Z/Z_{max}$. Thus, non-spherical assemblies are represented by bigger markers. Since we only focus on the crew-cut forming conditions, the solvent is modelled as a good solvent for the B block ($\chi_{BS}\leq 0.5$) in all the phase diagrams constructed.

For all the polymeric systems, there is a critical solvent-core interaction that defines the system's order-disorder transition (ODT) (see Figure~\ref{fig:blockLocalization}). In Figure~\ref{fig:PDexample}.b, the dashed line indicates this critical condition. We account for this ODT by defining an asymptotic phase parameter written as
\begin{equation}
\mathcal{F}_{ODT} = \frac{\chi_{\theta}N}{\chi_{AS}N}, 
\end{equation} 
where $\chi_{\theta}$ is the Flory-Huggins interaction parameter at theta condition. Following the traditional approach of Flory, $\chi_{\theta} = 0.5$. $\mathcal{F}_{ODT}$ is equal or larger than zero for $A$-solvent interactions below the ODT, and $\mathcal{F}_{ODT}<0$ when the segregation of the $A$ block occurs leading to the assemblies formation. Thus, micelle-forming solutions must have negative values of $\mathcal{F}_{ODT}$. 

In Figure~\ref{fig:PDexample}.b, we highlight the existence of two regimes in the phase diagrams, the first corresponds to the condition $\chi_{AS} > \chi_{AB}$ where large assemblies emerge, whereas smaller micelles dominate when $\chi_{AS} < \chi_{AB}$. The enthalpic interactions $\chi_{AS}$ and $\chi_{AB}$ accounts for this transition in the phase diagram through
\begin{equation}
\mathcal{F}_{contact} = \frac{\chi_{AS}f_A N - \chi_{AB}N}{\chi_{\theta}N}. 
\end{equation} 	
We interpret $\mathcal{F}_{contact}$ as a selectivity parameter for the core-forming block. Above the ODT, the selectivity of the $A$ block by the solvent and the $B$ block governs the characteristics of the assembly. If the affinity between the solvent and the $A$ block is lower than the affinity between $A$ and $B$, aggregates with larger aggregation numbers emerge to minimize the interactions with the solvent. This strong core segregation entails an entropic penalty associated with stretching the $A$ block. On the other hand, as the $A$-$B$ affinity decreases ($\chi_{AS} < \chi_{AB}$), aggregates with lower aggregation numbers can be stabilized. Merging the micelles requires the local increment of contacts between $A$ and $B$; thus, the growth is unfavorable. Besides, the relatively weak segregation between the solvent and the core-forming blocks allows a greater degree of core swelling, reducing the core stretch.  Values of $\mathcal{F}_{contact}<0$ must be characteristic for solutions with monodisperse micelles.

Apart from spherical micelles, different shapes of the aggregates are possible in monodisperse conditions ($\mathcal{F}_{contact}<0$). The curvature of the core-corona interface determines the experimental conditions where spherical micelles emerge. In weak-segregation conditions for the $A$ block, the repulsion between the $A$ and $B$ blocks governs the energy of the interface. Thus, a weak repulsion between $A$ and $B$ creates a diffuse interface with relatively small bending energy and promotes the formation of non-spherical assemblies. In contrast, as the core-corona repulsion increases, the interfaces become sharper, and small spherical micelles are stabilized. In short, $\mathcal{F}_{contact}$ accounts for the solution's monodispersity. We also introduce a phase parameter $\mathcal{F}_{corona-core}$ to consider the effect of the blocks' interfacial interactions, through 
\begin{equation}
\mathcal{F}_{corona-core} = \frac{\chi_{\theta}N - \chi_{AB}N}{\chi_{\theta}N}.
\end{equation} 	
Similarly to the previous phase parameters, we define the negative values of $\mathcal{F}_{corona-core}$ as distinctive for spherical aggregates. Weak interactions between blocks associated with elongated micelles near the theta point lead to $\mathcal{F}_{corona-core} \approx 0$.

\begin{figure}[!htb]
    \centering
    \includegraphics[trim=0cm 0cm 0cm 0cm, clip=true,width=0.9\textwidth]{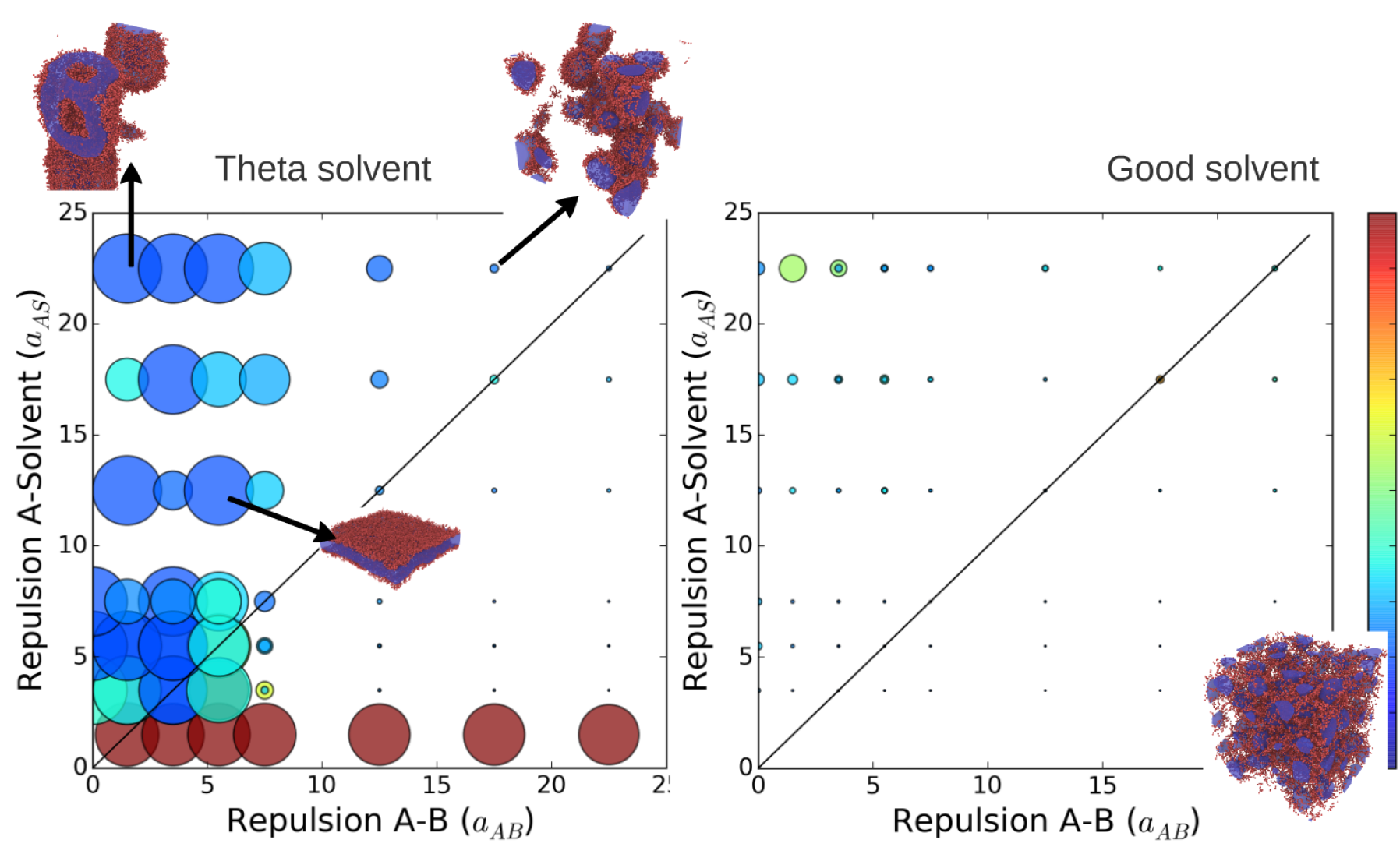}
    \caption[Corona swelling effect on micelle morphology]{Effect of the corona swelling over the size distribution and shape of the assemblies. Phase diagrams indicating polydispersity (colormap) and shape (size of the marker) of the aggregates for a) theta and b) good solvent conditions for the B block. A blocks are depicted as a solid surface, B blocks use particle representations, and solvent beads are omitted to facilitate the visualization. Theta solvents for the corona-forming block induce a variety of morphologies, including lamella and vesicles. Good solvents for the B block favor the swelling of the corona with solvent, favoring stable crew-cut morphologies.}
	\label{fig:PDsolventB}
\end{figure}

\subsubsection{Corona-Solvent Interaction Effect}

Figure~\ref{fig:blockLocalization} shows that both good and nearly theta solvents for the $B$ block induce crew-cut morphologies. However, these crew-cut structures adopt different shapes depending on the degree of swelling of the corona. Favorable interactions with the $B$ segments increase the swelling in good solvents. Figure~\ref{fig:PDsolventB} shows the variations in the phase diagram for block copolymers in good and theta solvent for the $B$ block. The regions of spherical micelle stability are enhanced as the solvent is highly selective for the corona, whereas, for theta conditions, the amount of stable morphologies increases (lamella, sphere, vesicles). 

In homopolymers, strong solvent segregation leads to large aggregates that minimize the number of contacts with the surrounding solvent. We identify a similar trend for block copolymer that increases the aggregation number when $\chi_{AS}>0.5$. However, the balance of the corona-solvent interactions is important in stabilizing and limiting aggregate size growth. In good solvents for the $B$ block, the larger degree of swelling of the corona preserves the core of the aggregates and restricts the growth of the micelles. The selective localization of the solvent in the corona avoids the encounter of the formed cores, imposing an entropic penalty for the micelle merging. On the other hand, as the amount of solvent in the corona region decreases (the solvent is less selective for $B$), the interfacial energy between the $A$ and $B$-rich domains diminishes facilitating the merging of the micelles and the formation of double-layer aggregates such as vesicles and films. This reduction in the interfacial energy is only compensated if the repulsion between the blocks is large. In such conditions, the effective interfacial tension between the core and corona is dominated by the block segregation, and the assemblies exhibit lower aggregation numbers, as Figure~\ref{fig:PDexample}.b shows. 

\begin{figure}[h!]
\centering
\includegraphics[width=0.48\textwidth]{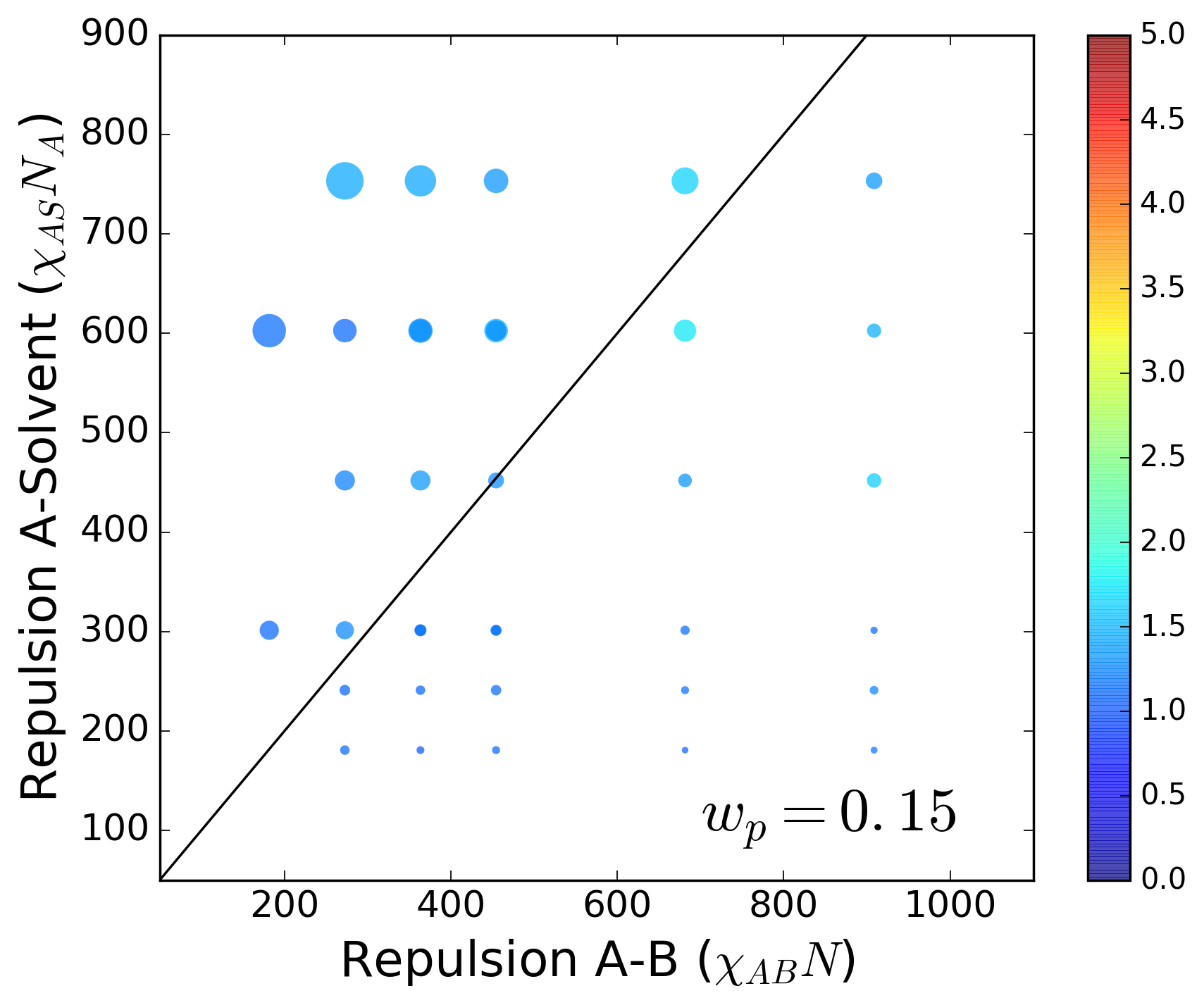}
\includegraphics[width=0.48\textwidth]{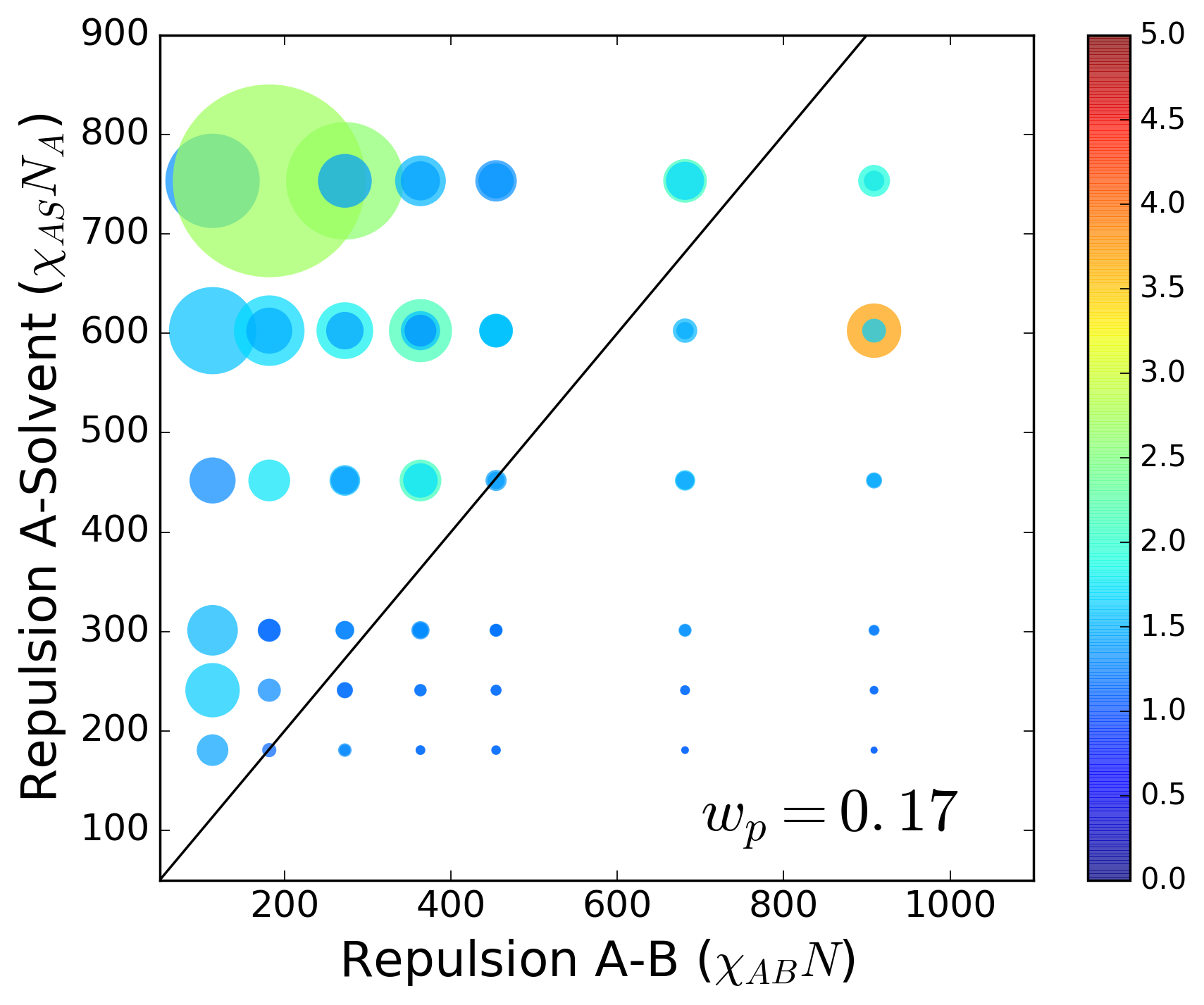}
\\
\includegraphics[width=0.48\textwidth]{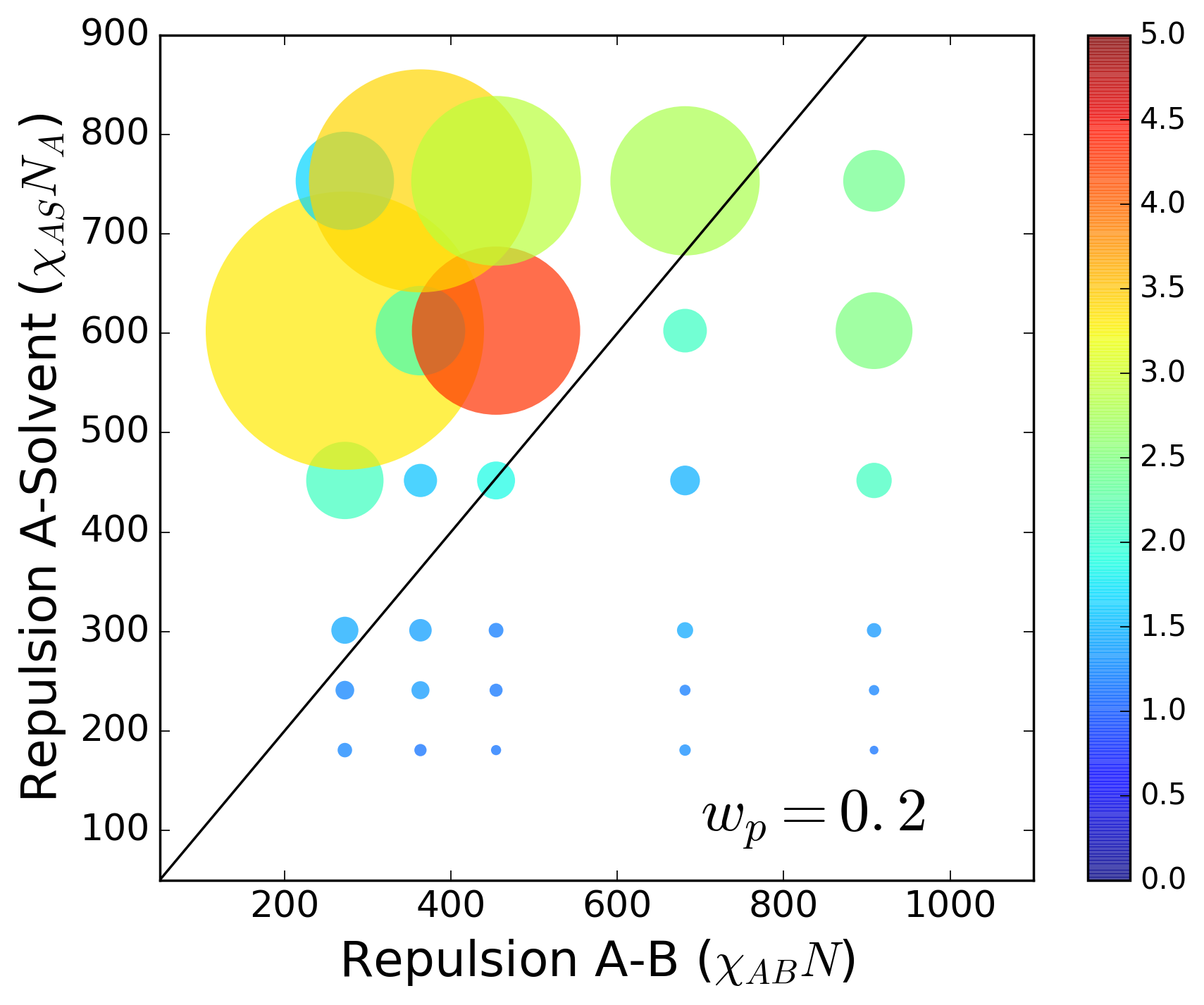}
\includegraphics[width=0.48\textwidth]{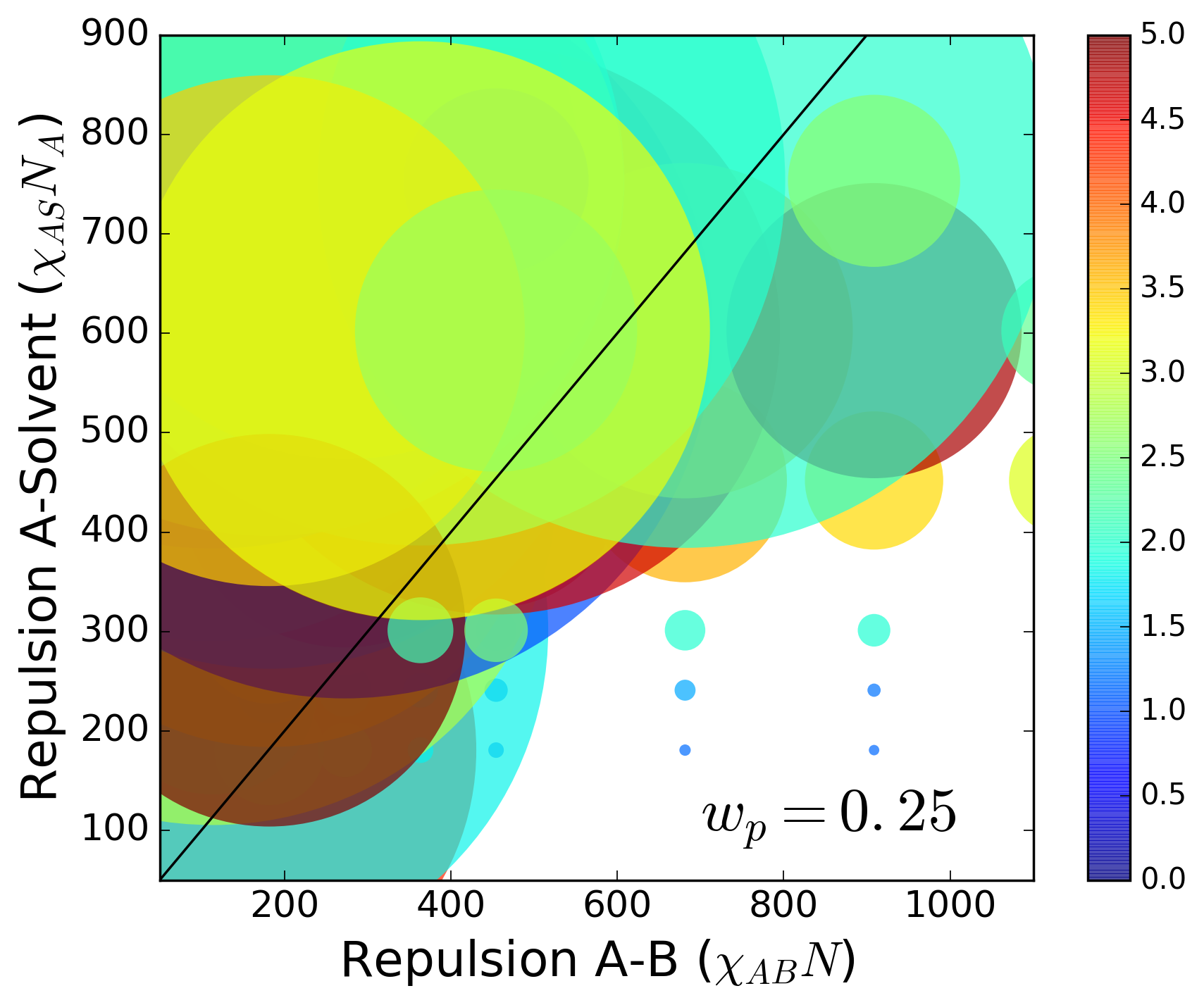}
\caption[Effect of the polymer concentration over the size distribution and shape of the assemblies]{Effect of the polymer concentration over the size distribution and shape of the assemblies. The increase in concentration favors the formation of large assemblies in regions of the phase diagram with strong segregation of the core-forming block. In contrast, weakly segregated cores are more likely to preserve the homogeneity of the assemblies as the concentration increase.}
\label{fig:PDconcentration}
\end{figure}
In summary, we identify that highly selective solvents for the corona facilitate the formation of monodisperse and spherical micelles. We account for this effect in terms of the $\chi_{BS}$ as
\begin{equation}
\mathcal{F}_{corona-solvent} = \frac{\chi_{BS}(1-f_A)N - \chi_{\theta}(1-f_A)N}{\chi_{\theta}N}.
\end{equation}     
The parameter $\mathcal{F}_{corona-solvent}$ takes positive values if the solvent is poor for the $B$ block, and negative for good solvents. Therefore, $\mathcal{F}_{corona-solvent}<0$ characterizes systems with monodisperse and spherical micelles.

\subsubsection{Polymer Concentration Effect}

So far, we identify the energetic conditions between solvent and block copolymers that ensure the formation of monodisperse-spherical micelles in solution and postulate a set of parameters that characterize these  favorable conditions. We now evaluate the polymer concentration's effect on the micelles' ordering. Polymer concentration effects are related to the micellar volumetric fraction. The polymer concentration in casting solutions typically ranges between 15\% to 25\%, depending on the polymer-solvent solution used. Figure~\ref{fig:PDconcentration} presents the phase diagram for four different polymer concentrations (15\%, 17\%, 20\%, and 25\%). The increase in the polymer concentration, in general, promotes the growth of the aggregation number of the micelles in regions of the phase diagram with strong segregation of the core-forming block. In systems with weakly segregated cores, because of the monodispersity of the micelles the system preserves its homogeneity as the concentration increase.

\begin{figure}[h!]
	\centering
	\begin{minipage}{.48\textwidth}
		\centering
		\includegraphics[trim=0cm 0.7cm 0cm 0cm, clip=true,width=1\linewidth]{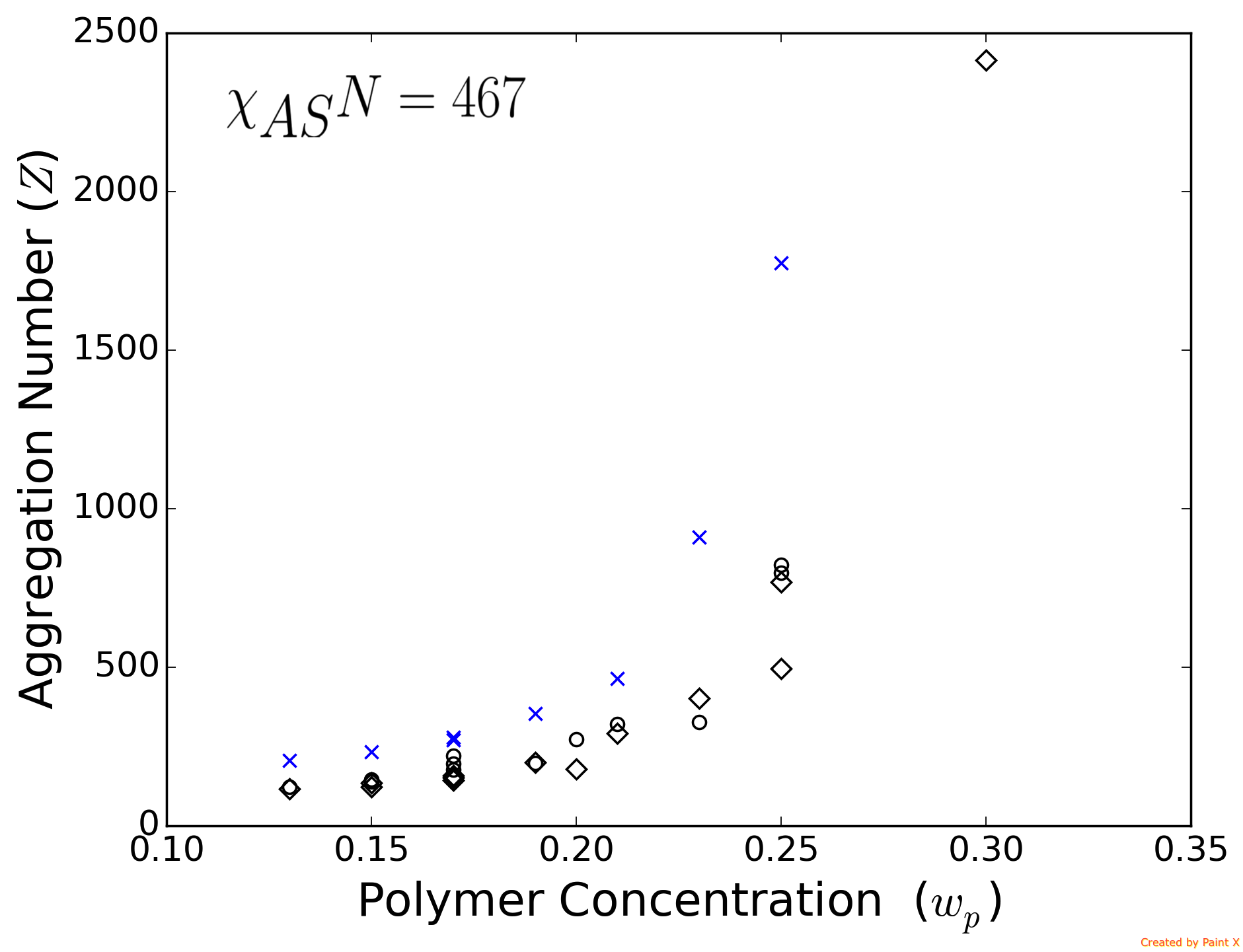}
		\label{fig:prob1_6_2}
	\end{minipage}
	\begin{minipage}{0.48\textwidth}
		\centering
		\includegraphics[trim=0cm 0.7cm 0cm 0cm, clip=true,width=1\linewidth]{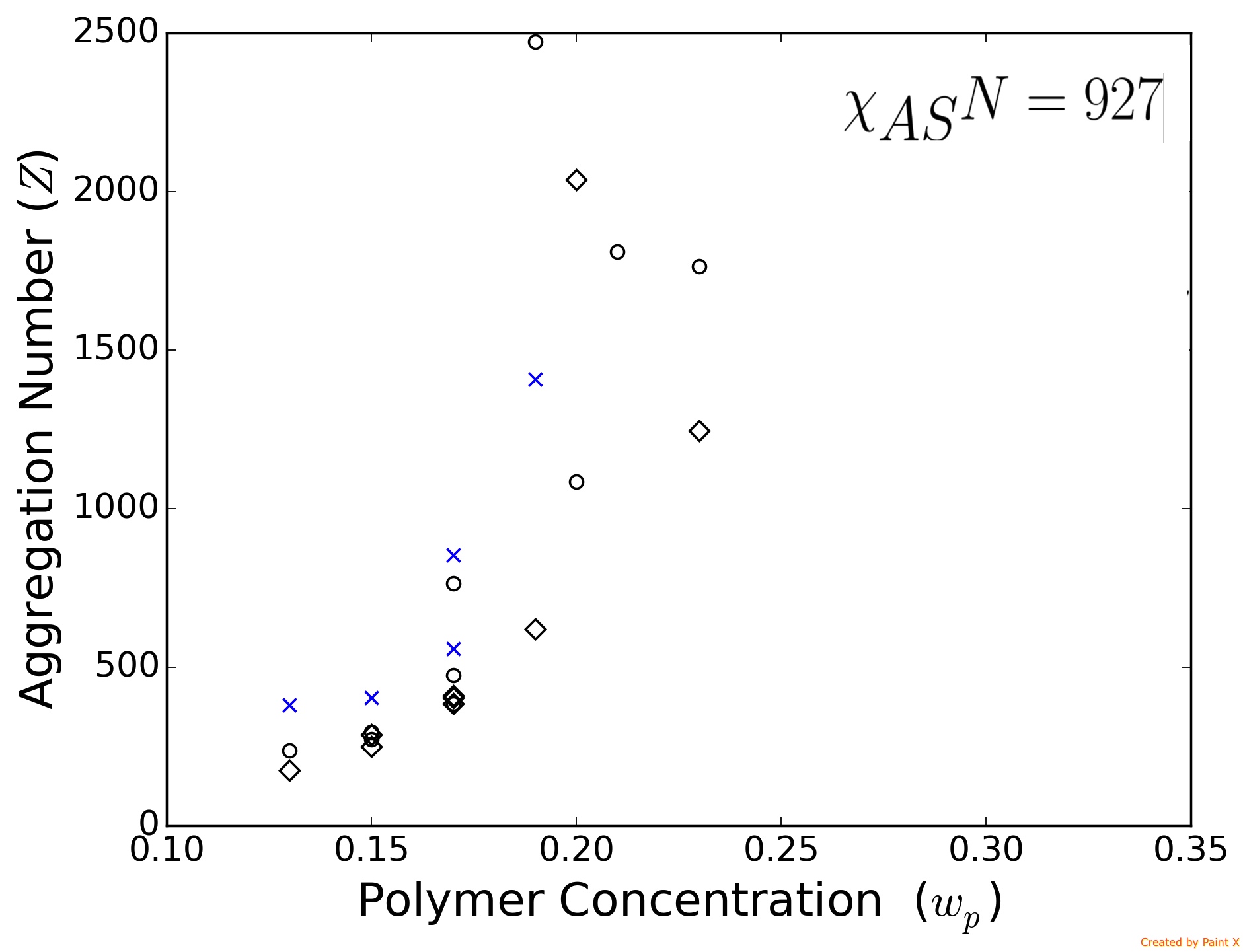}
		
		\label{fig:prob1_6_1}
	\end{minipage}
\caption[Variation of the aggregation number with the polymer concentration]{Variation of the aggregation number ($Z$) with the polymer concentration($w_p$) for two different A-solvent interactions ($\chi_{AS}$), and different segregation strengths between blocks: ({x}) $\chi_{AB}N=371$; ($\circ$) $\chi_{AB}N=464$; ($\diamond$) $\chi_{AB}N=695$. Aggregation conditions with spherical and monodisperse micelles exhibit a slight increase in the micellar size with the increment in the polymer concentration up to a certain concentration threshold where large cluster formation occurs. This threshold is much smaller in polydisperse systems, where large aggregates are favored due to the strong segregation of the core-forming block.}
\label{fig:concEffectInZ}
\end{figure}

In solutions with monodisperse-spherical micelles, the increase in the concentration modifies the packing of the micelles but does not affect their size if the volumetric concentration is smaller than $v_{max}$. Figure~\ref{fig:PDconcentration} shows that the increment in the polymer concentration in regions of the phase diagram where the micelles are small and well distributed does not significantly modify the shape and aggregation number of the micelles. On the contrary, larger clusters appear as the polymer concentration rises in regions of polydisperse micelles. Figure~\ref{fig:concEffectInZ} presents the variation of $Z$ with the polymer concentration for two different core segregation conditions ($\chi_{AS}$) varying the segregation strengths between blocks ($\chi_{AB}$). Systems that form spherical and monodisperse micelles only exhibit a slight increase in the micellar size with the increment in the polymer concentration up to a certain threshold, where large cluster formation occurs. We identify this critical concentration with $w_{max}$ before the micelle overlapping happens. In Figure~\ref{fig:concEffectInZ}, the threshold is much smaller in systems with strong core segregation (polydisperse systems), where large aggregates are favored. 


We introduce an entropic phase parameter $\mathcal{F}_{pack}$ that accounts for the packing condition of the micelles. We construct this parameter such that $\mathcal{F}_{pack}<0$ when the volumetric fraction of micelles approximates to the close-packing condition. $\mathcal{F}_{pack}$ is given by
\begin{equation}
\mathcal{F}_{pack} = \bar{w}_M\text{ln}\bar{w}_M + (1-\bar{w}_M)\text{ln}(1-\bar{w}_M),
\label{eq:fpacking}
\end{equation}
where $\bar{w}_M = w_M - (1-w_{max})$. According to~\eqref{eq:fpacking}, the minimum of $\mathcal{F}_{pack}$  occurs when the volumetric fraction of the micelles is equal to the maximum value for rigid-sphere close packing.

\begin{figure}[!htb]
\centering
    \begin{minipage}{.33\textwidth}
        \centering
        \includegraphics[trim=0cm 0.0cm 0cm 0cm, clip=true,width=1\linewidth]{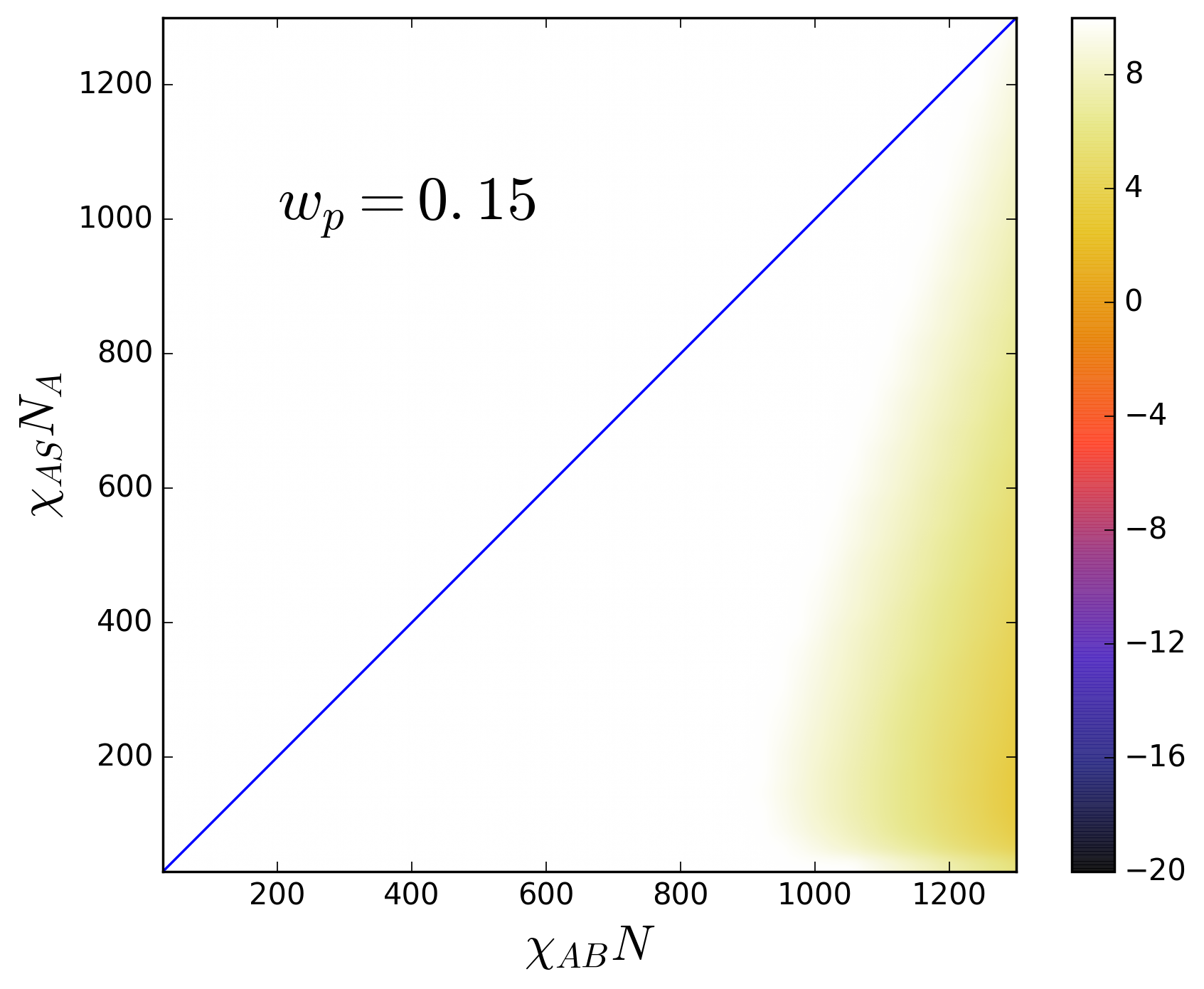}
    \end{minipage}
    \begin{minipage}{0.33\textwidth}
        \centering
        \includegraphics[trim=0cm 0.0cm 0cm 0cm, clip=true,width=1\linewidth]{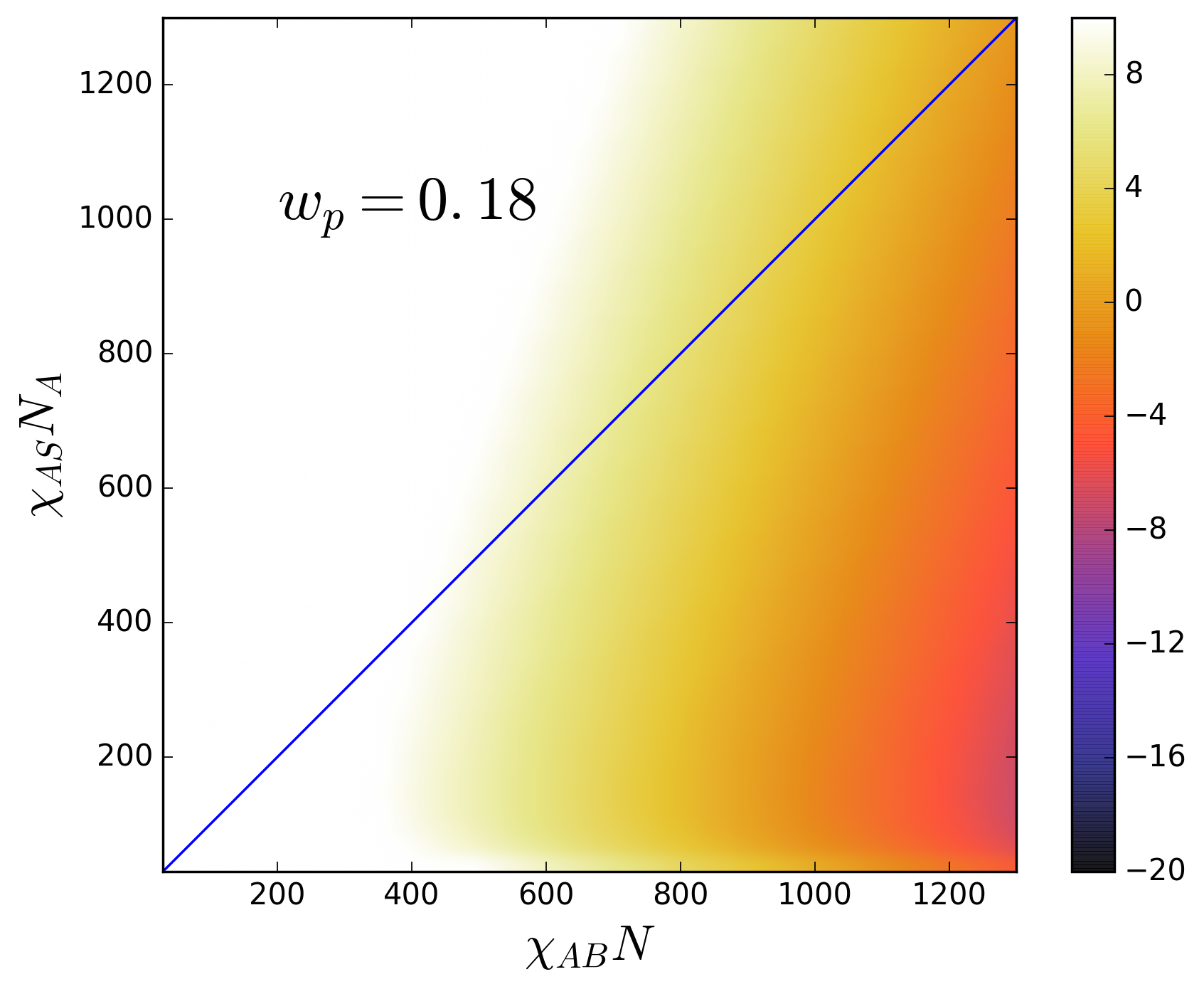}
    \end{minipage}
    \begin{minipage}{0.33\textwidth}
        \centering
        \includegraphics[trim=0cm 0.0cm 0cm 0cm, clip=true,width=1\linewidth]{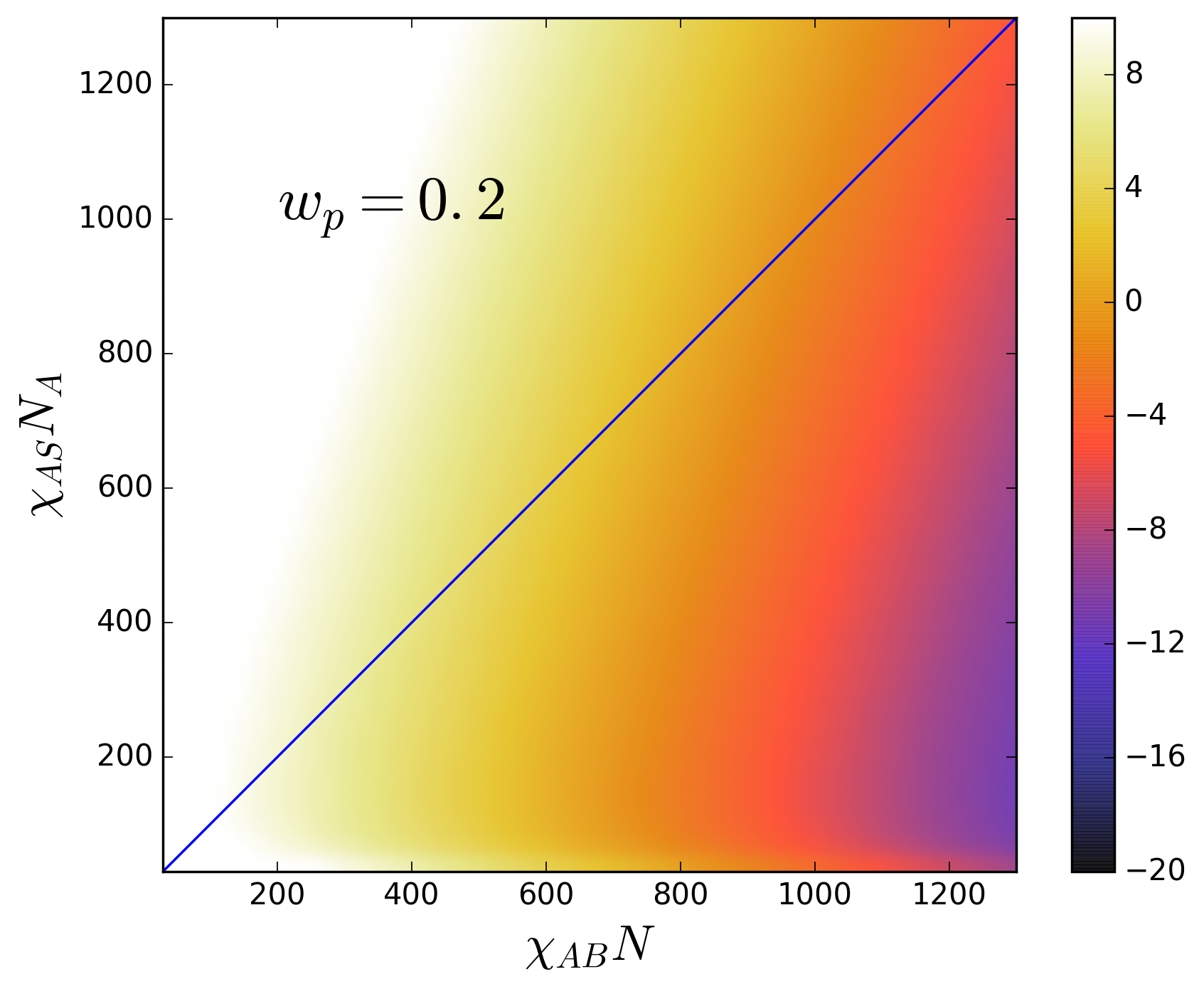}
    \end{minipage}
    
    \begin{minipage}{0.33\textwidth}
        \centering
        \includegraphics[trim=0cm 0.0cm 0cm 0cm, clip=true,width=1\linewidth]{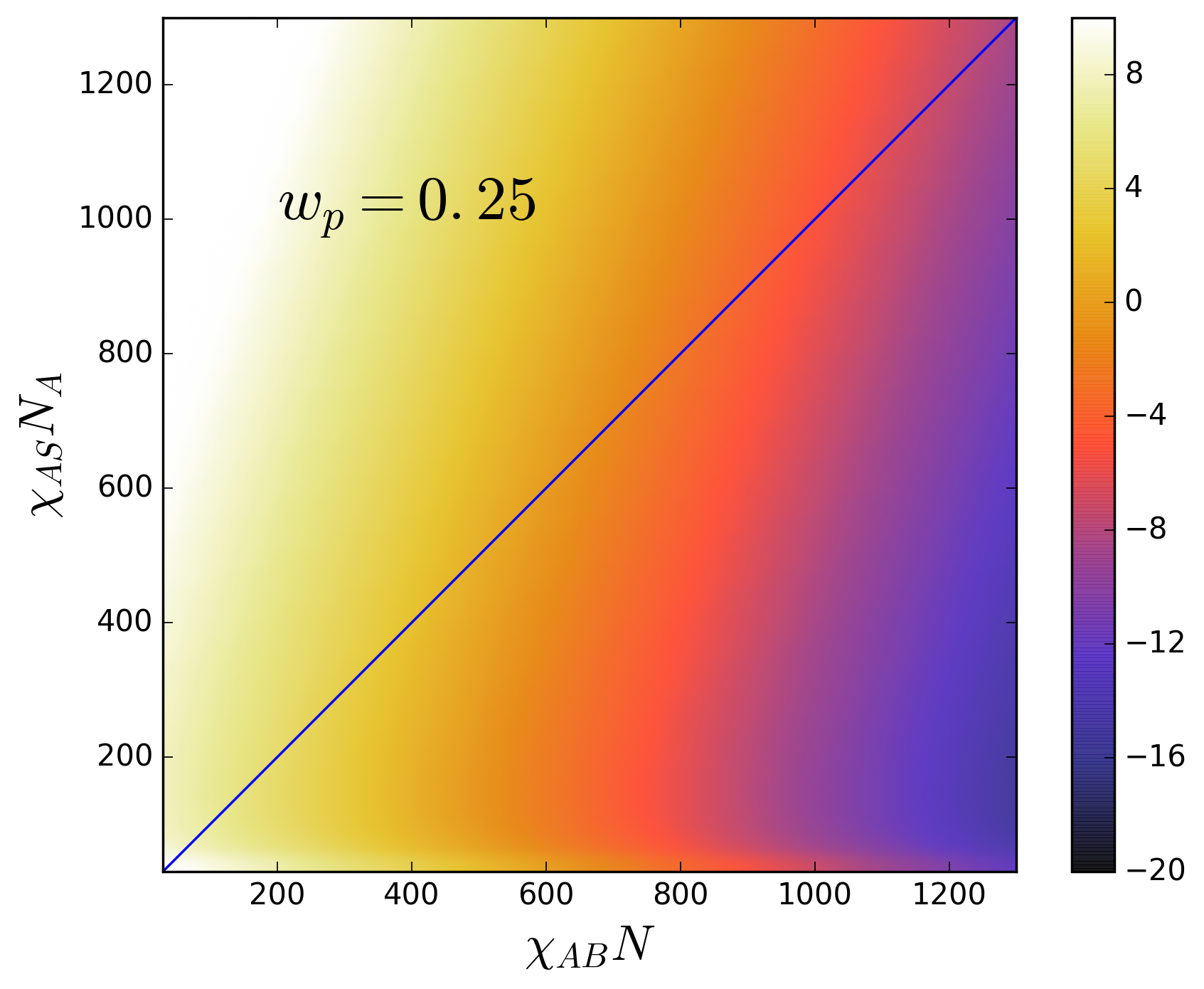}
    \end{minipage}
    \begin{minipage}{0.33\textwidth}
        \centering
        \includegraphics[trim=0cm 0.0cm 0cm 0cm, clip=true,width=1\linewidth]{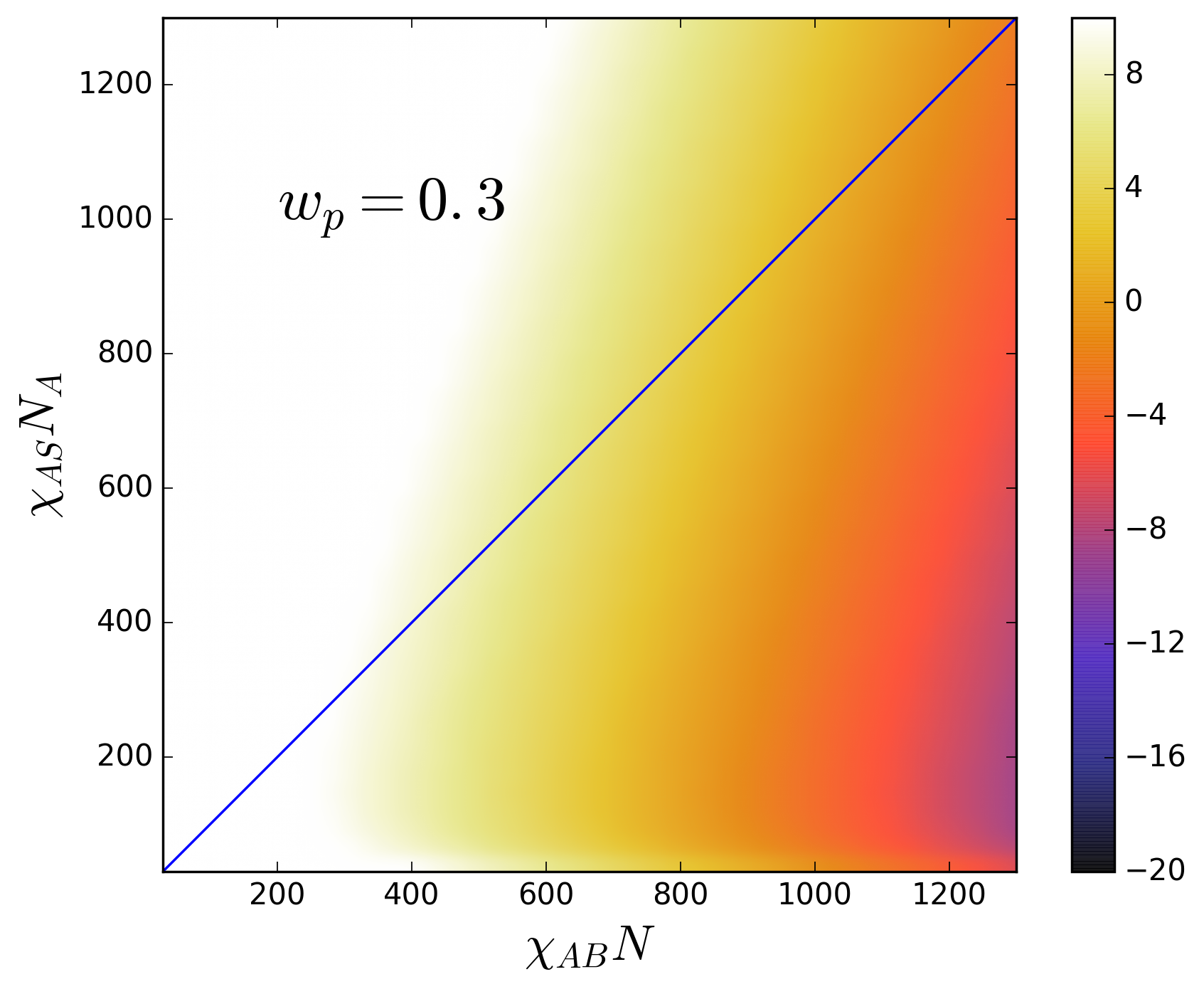}
    \end{minipage}%
    \begin{minipage}{0.33\textwidth}
        \centering
        \includegraphics[trim=0cm 0.0cm 0cm 0cm, clip=true,width=1\linewidth]{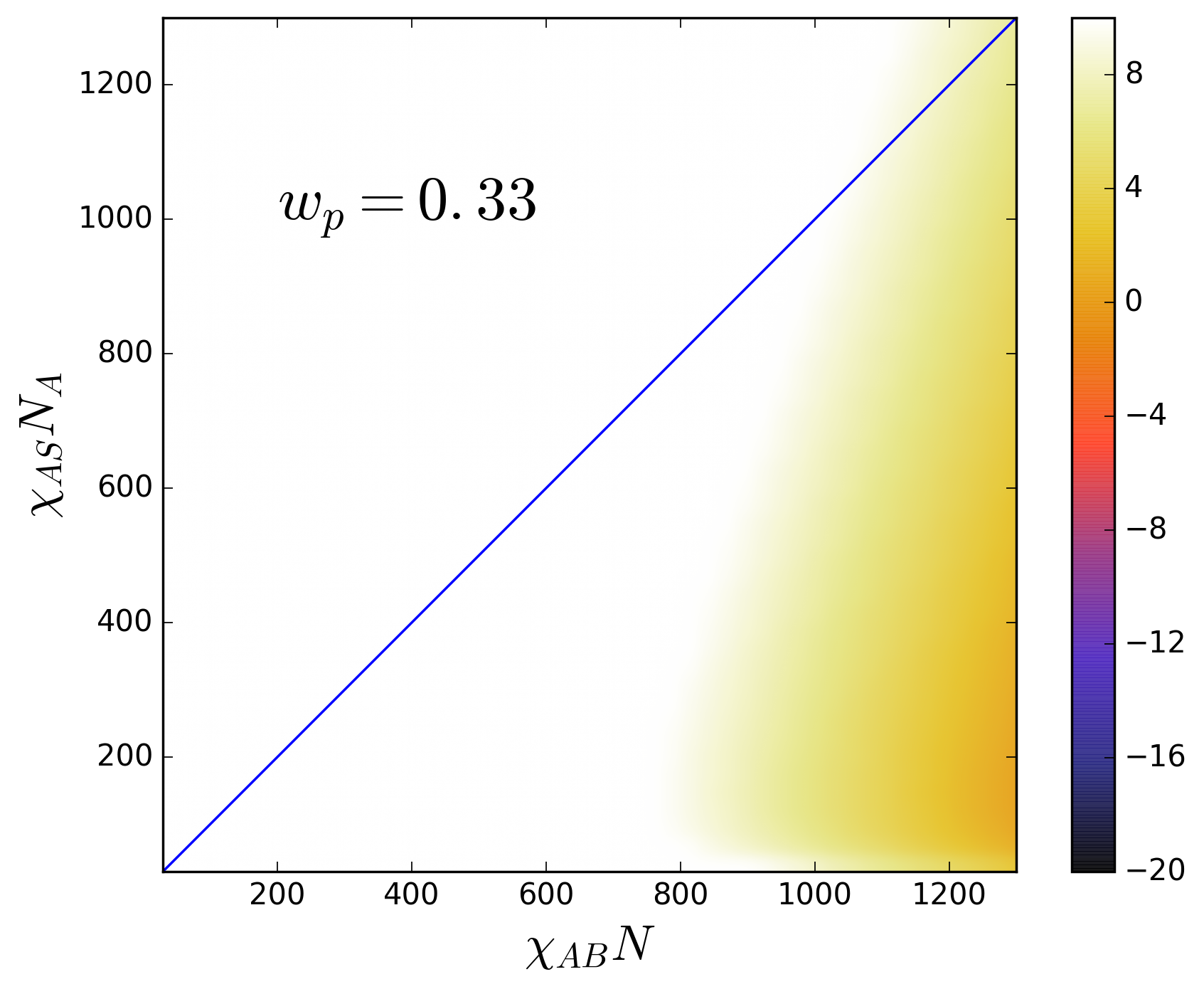}
    \end{minipage}	
\caption[Global phase parameter variation $\mathcal{F}$]{Global phase parameter variation $\mathcal{F}$ as a function of $\chi_{AB}N$ and $\chi_{AS}N_A$ for solutions with different polymer concentrations}
\label{fig:fitting}
\end{figure}

\subsection{Phase-Diagram Generalization}

The balance between entropic and enthalpic interactions between the blocks and the solvent defines the phase equilibria of the micelles. We confirm these transitions in all the casting-solution simulations. Given the phase behaviour of the assembled block copolymer, we propose a generalized expression that describes the morphological transitions we present. Thus, we introduce a global phase parameter $\mathcal{F}$ to describe the changes in the morphology of the aggregates as a contribution of all the phase parameters we introduced above. Thus, $\mathcal{F}$ is a phenomenological parameter that decreases as the micellar system becomes spherical, monodisperse, and closely packed. We compute the magnitude of $\mathcal{F}$ as
\begin{equation}
\mathcal{F} = \mathcal{F}_{contact} + \mathcal{F}_{corona-solvent} + \mathcal{F}_{corona-core} + \mathcal{F}_{ODT} + K_{pack}\mathcal{F}_{pack}.
\label{eq:fglobal}
\end{equation}
where $K_{pack}=100$ is a constant that scales the effect of $\mathcal{F}_{pack}$ to the same order of magnitude of the other parameters. Negative values of $\mathcal{F}$ characterize the solutions containing monodisperse-spherical micelles. In equation~\eqref{eq:fglobal}, the first three terms account for the energy of the interface. Substituting each interfacial term with its definition, we rewrite $\mathcal{F}$ as
\begin{equation}
\mathcal{F} = \frac{(\chi_{AS}f_AN-\chi_{AB}N)+(\chi_{BS}(1-f_A)N-\chi_{AB}N)}{\chi_{\theta}N} + \mathcal{F}_{ODT} + K_{pack}\mathcal{F}_{pack}.
\label{eq:fglobal2}
\end{equation}
Figure~\ref{fig:fitting} presents the variation of the global phase parameter for different polymer-solvent interactions. This global phase parameter variation is consistent with the morphological transitions in the phase diagram. 
%

\section*{Conclusions}

In conclusion, we describe a geometrical analysis of rigid-sphere packing that effectively describes the self-assembly of diblock copolymers in solution. Our approach narrows the experimental conditions of concentration and polymer-solvent interactions where micellar ordering in solution is obtained. We identify that weak solvent segregation of the largest block plays a crucial role in stabilising monodisperse assemblies and ordering in solution. On the other hand, if the solvent affinity with the core-forming block induces strong phase separation, larger and polydisperse aggregates are produced to minimize the unfavorable enthalpic interaction between species. The curvature of the core-corona interface also plays an important role in determining the experimental conditions where spherical micelles emerge. The energy of the interface is governed by the repulsion between the $A$ and $B$ blocks, and a weak $A$-$B$ segregation promotes the formation of non-spherical assemblies with diffuse interfaces and relatively small bending energy.

The new global phase parameter can be used as an initial predictor for ordering in solution if the Flory-Huggins parameters of the species are known. However, due to the inherent simplification of the Flory-Huggins model and the rigid-sphere approximation of the micelles used to characterize the assembly, some deviations of the computed parameter may exist. The main application of this approach is to narrow the interaction parameter concentration space of casting solution preparation, which can significantly reduce the trial-and-error methodology used in membrane preparation, saving time and resources. Our study provides valuable insights into the multi-scale phenomena involved in the self-assembly of diblock copolymers in solution and may find applications in various fields beyond isoporous membranes.

\section*{Contribution statement}

N.M, S.N, and V.C conceived the study. The first draft of the manuscript was written by N.M, and all authors contributed to
the final version of the manuscript. N.M conducted the simulations, experiments and data processing. All the authors
discussed and analyzed the results. All authors approved the final version of the manuscript.

\section*{Data availability}
Data supporting the findings of this paper are available from the corresponding authors upon request.

\section*{Declaration of Competing Interest}

The authors declare that they have no known competing financial interests or personal relationships that could have appeared to influence the work reported in this paper.

\section*{Acknowledgement}

The authors acknowledge the support of Basque Government through the BERC 2022-2025 program and by the Ministry of Science and Innovation: BCAM Severo Ochoa accreditation CEX2021-001142-S / MICIN / AEI / 10.13039/501100011033. N.M acknowledges the support from the European Union’s Horizon 2020 under the Marie Sklodowska-Curie Individual Fellowships grant 101021893, with the acronym ViBRheo

\newpage
 \appendix

\section{Scaling functions to construct coarse models}

DPD parameters of fine-scale representations are scaled to construct coarser polymer representations. Table~\ref{tab:scalingsMoreno} gives a breakdown of the scaling functions used to construct the coarse-grained models of the block copolymers. The main feature of this methodology is the proper conservation of the chain conformation explicitly between the original fine scale ($\nu'$) and the coarse representation ($\bar{\nu}$). The use of bond-angle potential controls the conformation of the coarse scales entropically. 
 
\begin{table}[!h]
\centering
 \caption {Coarse parameters $A_{coarse}$ and scaling function $\psi(\phi)$ proposed for coarse graining of systems with chains. }
    \begin{tabular*}{0.5\textwidth}{@{\extracolsep{\fill}}c|c}
    \hline
   \textbf{$A_{coarse}$} \quad   &  \textbf{$\psi(\phi)$}     \quad      \\ \hline \hline                                                                                                                              
 $\bar{m}$                 & $\phi$                                  \\ \hline                                            
 $\bar{N_T}$             & $\phi^{-1}$                          \\ \hline                                               
 $\bar{r}_c $               & $N'^{\nu' - \bar{\nu}}\phi^{\bar{\nu}}$     \\ \hline                                 
 $\bar{v}    $               & $(N'^{(\nu' - \bar{\nu})}\phi^{\bar{\nu}})^3$         \\ \hline                              
 $\bar{\nu}  $             & $\frac{1}{3}~\frac{({3\nu' \text{ln}N' - \text{ln}\phi})}{({\text{ln}N' - \text{ln}\phi})}$ \\ \hline
 $\bar{\tau}  $           &  $N'^{(\nu' - \bar{\nu})}\phi^{\bar{\nu}}$      \\ \hline                                     
 $\bar{\epsilon}$        & $\phi$                                                        \\ \hline                      
 $\bar{a}_{ij} $             & $\phi^{1-\bar{\nu}} N'^{(\bar{\nu}-\nu')}$      \\ \hline
 $\bar{K_s} $                & $(N' +\phi-1)\phi^{1-2\bar{\nu}}N'^{2(\nu'-\bar{\nu})-1}$ \\ \hline 
 $\bar{\gamma}$         & $\phi^{1-\bar{\nu}} N'^{(\bar{\nu}-\nu')}$                                                        \\ \hline 
 $\bar{\sigma}  $          & $\phi^{1-\frac{\bar{\nu}}{2}} N'^{\frac{(\bar{\nu}-\nu')}{2}}$     \\ \hline
 $\bar{K_a} $                 &$(N'+\phi-2)\frac{\phi}{N'}$    \\ \hline
    \end{tabular*}
 \label{tab:scalingsMoreno}
\end{table}



\bibliographystyle{unsrt}
\bibliography{refs}







\end{document}